\documentclass[reprint,floatfix,notitlepage,nofootinbib,twocolumn,superscriptaddress,prb]{revtex4-1}

\usepackage{amsmath}
\usepackage{amssymb}
\usepackage{graphicx}
\usepackage[tight]{subfigure}
\usepackage{enumitem}
\usepackage{soul}
\usepackage{cancel}
\usepackage{multirow}
\usepackage{tikz}
\usepackage{pifont}
\usepackage{xspace}
\usepackage{comment}

\makeatletter
\renewcommand{\p@subsection}{}
\renewcommand{\p@subsubsection}{}
\makeatother

\usepackage[english]{babel}
\makeatletter
\def\bbl@set@language#1{%
  \edef\languagename{%
    \ifnum\escapechar=\expandafter`\string#1\@empty
    \else\string#1\@empty\fi}%
  \@ifundefined{babel@language@alias@\languagename}{}{%
    \edef\languagename{\@nameuse{babel@language@alias@\languagename}}%
  }%
  \select@language{\languagename}%
  \expandafter\ifx\csname date\languagename\endcsname\relax\else
    \if@filesw
      \protected@write\@auxout{}{\string\select@language{\languagename}}%
      \bbl@for\bbl@tempa\BabelContentsFiles{%
        \addtocontents{\bbl@tempa}{\xstring\select@language{\languagename}}}%
      \bbl@usehooks{write}{}%
    \fi
  \fi}
\newcommand{\DeclareLanguageAlias}[2]{%
  \global\@namedef{babel@language@alias@#1}{#2}%
}
\makeatother
\DeclareLanguageAlias{en}{english}

\usetikzlibrary{arrows}
\usetikzlibrary{intersections}
\usetikzlibrary{shapes.geometric}
\usetikzlibrary{decorations.pathmorphing, patterns,shapes,fixedpointarithmetic}
\usetikzlibrary{decorations.markings}

\tikzset{
  mid arrow/.style={postaction={decorate,decoration={
        markings,
        mark=at position .575 with {\arrow{stealth}}
      }}},
  end arrow/.style={postaction={decorate,decoration={
        markings,
        mark=at position 1 with {\arrow{stealth}}
      }}},
  snake arrow/.style={fixed point arithmetic, decorate, decoration={snake,amplitude=2pt, segment length=11pt},postaction={decoration={markings,mark=at position 0.625 with {\arrow{stealth}}},decorate}},
}

\usepackage{bm}

\usepackage[pdfusetitle,colorlinks=true,citecolor=blue,linkcolor=magenta]{hyperref}

\graphicspath{{./}{./images/}}

\begin{document}
\title{Measurement-induced criticality in \texorpdfstring{$\mathbb{Z}_2$}{}-symmetric quantum automaton circuits}

\author{Yiqiu Han}
\email{hankq@bc.edu}
\affiliation{Department of Physics, Boston College, Chestnut Hill, MA 02467, USA}

\author{Xiao Chen}
\email{chenaad@bc.edu}
\affiliation{Department of Physics, Boston College, Chestnut Hill, MA 02467, USA}

\begin{abstract}
We study entanglement dynamics in hybrid $\mathbb{Z}_2$-symmetric quantum automaton circuits subject to local composite measurements. We show that there exists an entanglement phase transition from a volume-law phase to a  critical phase by varying the measurement rate $p$. By analyzing the underlying classical bit-string dynamics, we demonstrate that the critical point belongs to parity-conserving universality class. We further show that the critical phase with $p>p_c$ is related to the diffusion-annihilation process and is protected by the $\mathbb{Z}_2$-symmetric measurement. We give an interpretation of the entanglement entropy in terms of a two-species particle model and identify the coefficient in front of the critical logarithmic entanglement scaling as the local persistent coefficient. The critical behavior observed at $p\geq p_c$ and the associated dynamical exponents are also confirmed in the purification dynamics.
\end{abstract}
\maketitle

\section{Introduction}
Precisely manipulating qubits and mitigating noise have become key tasks in the noisy intermediate-scale quantum (NISQ) era. Recently, it has been shown that monitoring many-body quantum systems with active measurements can induce a quantum information phase transition\cite{skinner2019measurement,Li_2018,Chan_2019}. When the monitoring frequency is small, the information of the system is protected by the unitary evolution and the wave function is still a highly entangled volume-law state. As the monitoring frequency is increased, the unitary evolution cannot effectively protect the quantum information, and the system undergoes a phase transition to a disentangled area-law state. 

This phase transition was first observed in Haar random and Clifford random circuits composed of local two-qubit unitary gates and single qubit projective measurement gates\cite{Nahum_2017,skinner2019measurement,Li_2018,Chan_2019,choi2020quantum,Gullans_2020,Li_2019}. In these quantum circuits, increasing the measurement rate leads to an entanglement phase transition from a volume-law phase to an area-law phase if we follow the quantum trajectories. In particular, at the phase transition point aspects of critical phenomena come into play, with, e.g., emergent two-dimensional conformal symmetry emerging in certain (1+1)-dimensional [(1+1)D] circuits\cite{Nahum_2017,li2020conformal}.  Since its discovery, this phase transition has been generalized to other monitored open quantum dynamics\cite{Tang_2020,Szyniszewski_2019}. It has an interesting interpretation in terms of quantum error correction\cite{choi2020quantum,Gullans_2020} and can be understood as a symmetry-breaking phase transition in the enlarged replica space\cite{jian2019measurementinduced,bao2020theory,Jian_2021,bao2021symmetry}, where the entanglement entropy corresponds to the domain wall free energy. Recently, it was shown in Ref.~\onlinecite{Iaconis:2021pmf} that the quantum automaton (QA) circuit subject to composite measurement can also exhibit an entanglement phase transition. This model provides a new physical picture for interpreting the phase transition in terms of bit-string dynamics and the entanglement transition within this model belongs to the directed percolation (DP) universality class\cite{henkel2008non}. 

Monitoring quantum systems can also stabilize interesting phases which cannot exist in equilibrium. One example is non-unitary free-fermion dynamics. In this system, there is an emergent critical phase protected by continuous weak measurement\cite{Chen_2020,alberton2020trajectory}. Another class of examples are given by monitored quantum systems with additional discrete symmetries, which can possess highly entangled volume-law phases with conventional or topological order\cite{Sang_2021,Lavasani_2021,bao2021symmetry}. In addition, the area-law phase can also have a richer phase diagram characterized by different orders\cite{Sang_2021,Lavasani_2021,Ippoliti_2021}.

\begin{figure}
\centering

  \includegraphics[scale=0.42]{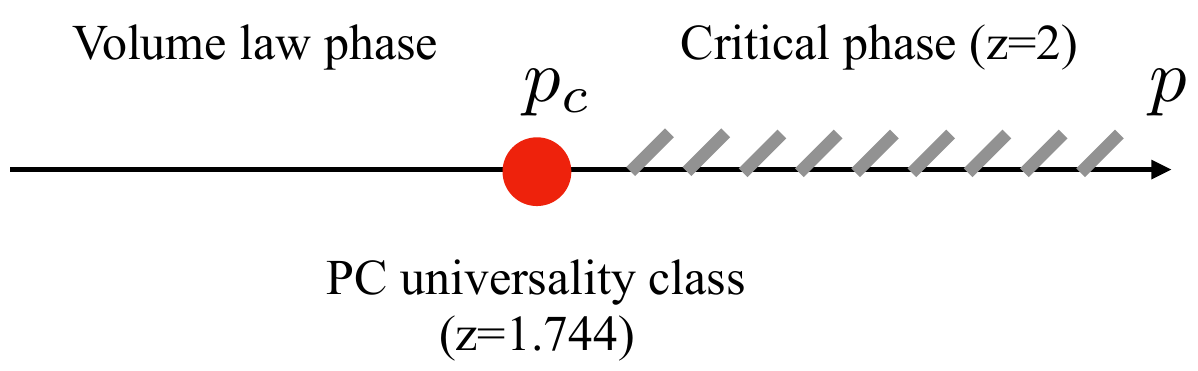}
\caption{A cartoon picture for the phase diagram of the hybrid QA circuit in the presence of $\mathbb{Z}_2$ symmetry. The dynamical exponents of the quantum phase transition at $p=p_c$ and the quantum critical phase $p>p_c$ are inherited from the associated classical bit-string dynamics, respectively.} 
\label{fig:schematics}
\end{figure}

Motivated by the above works, in this paper we construct a hybrid QA circuit with $\mathbb{Z}_2$ symmetry and study its entanglement dynamics. We show that if we impose this $\mathbb{Z}_2$ symmetry, there exists an entanglement phase transition from a highly entangled volume-law phase to a critical phase with logarithmic entanglement scaling, with the transition occurring by varying the measurement rate $p$ (See Fig.~\ref{fig:schematics}). We generalize the classical bit-string picture developed in Ref.~\onlinecite{Iaconis:2021pmf} and demonstrate that the entanglement phase transition belongs to the parity-conserving (PC) universality class with dynamical exponent $z=1.744$\cite{ZHONG1995333,Park_2001,henkel2008non}. Due to the $\mathbb{Z}_2$ symmetry, this universality class is distinct from the aforementioned DP universality class. We further derive a two-species particle model based on the bit-string picture to calculate the entanglement dynamics from a short-range entangled state. The particles in this model can diffuse, branch, and annihilate in pairs, and the purity for a subsystem is equivalent to the fraction of configurations where particles of different species never encounter one another. In particular, the prefactor of the logarithmic scaling of the second R\'enyi entropy at the transition point $p_c$ is related to the local persistence coefficient of the two-species particle model and is a universal constant for PC universality class. 

Unlike the conventional measurement-induced phase transition in which there is an area-law entangled phase when the measurement rate $p$ is larger than some critical threshold $p_c$, here we observe a critical {\it phase}, characterized by logarithmic entanglement scaling when $p>p_c$. Specifically, this phase has dynamical exponent $z=2$ and is protected by the combination of the $\mathbb{Z}_2$ symmetry and the special features of the QA circuit.
We show that the underlying bit-strings have diffusive dynamics, and provide an interpretation of the critical entanglement scaling in terms of the two-species particle model. We further analyze the purification dynamics starting from a mixed density matrix with extensive entropy\cite{Gullans_2020}. We find that when $p>p_c$, the entropy decays diffusively in time which is consistent with the entanglement dynamics results.  

The rest of the paper is organized as follows. In Sec.~\ref{sec:QA}, we construct a hybrid QA circuit with $\mathbb{Z}_2$ symmetry. We numerically compute the entanglement entropy for this circuit in Sec.~\ref{Sec: EE} in terms of a Clifford stabilizer representation. In addition, we provide an interpretation of second R\'enyi entropy in terms of classical particle model. In Sec.~\ref{Sec: PT}, we analyze the purification dynamics and find that the results for critical point and critical phase are consistent with that in Sec.~\ref{Sec: EE}. We summarize our results in Sec.~\ref{sec:concl}.

\section{QA model with \texorpdfstring{$\mathbb{Z}_2$}{} symmetry}\label{sec:QA}
In this section, we construct a hybrid QA circuit with $\mathbb{Z}_2$ symmetry. We aim to study how the information encoded in the quantum state evolves under the competition between quantum automaton unitary operators and non-unitary measurements, which will be specified later in this section. Given a subregion $A$, a particularly useful quantity to measure this is the $n^{th}$ R\'enyi entropy:
\begin{equation}
\begin{aligned}
  S^{(n)}_A &=\frac{1}{1-n}\ln{[\text{Tr}(\rho_A^n)]}
\\
  \rho_A &=\text{Tr}_B|\psi\rangle\langle\psi|.
  \end{aligned}
\end{equation}
where $B$ is the complement of $A$. In this paper, we will focus on the second R\'enyi entropy  with $n=2$  and take the base to be the natural logarithm base. 

The QA circuit is built up of unitary operators that permute a set of vectors in a specific orthonormal basis (namely, the computational basis) up to some random phase, i.e.,
\begin{equation}
  U|n\rangle=e^{i\theta_n}|\pi(n)\rangle,
\end{equation}
where $\pi\in S_{N}$ is an element of the permutation group on the product states $|n\rangle$ in the computational basis with cardinality $N$. Through out this paper, we build the computational basis from the Pauli $Z$ basis. The $\mathbb{Z}_2$ symmetry is imposed by requiring that the parity of the computational basis remains fixed under the unitary evolution. From the previous definition it is clear that the automaton unitary evolution does not create entanglement when acting on product states in the computational basis. However, it can generate entanglement in a wavefunction which involves a superposition of the basis states---for example, we can apply the measurement $(1+Z_1Z_2\cdots Z_L)/\sqrt{2}$ to a product state polarized in $+x$ direction with an even number of qubits $L$ to make it $\mathbb{Z}_2$ even. When the automaton unitary operator acts on such an initial state,
\begin{equation}
  \begin{aligned}
  |\psi_I\rangle=U|\psi_0\rangle &=U\circ \frac{1+Z_1Z_2\cdots Z_L}{\sqrt{2}}\bigotimes_i\frac{1}{\sqrt{2}}(|0\rangle +|1\rangle)
  \\
  &=\frac{1}{\sqrt{2^{L-1}}}\sum_n e^{i\theta_n}|\pi(n)\rangle,
  \end{aligned}
\end{equation}
we can obtain a highly entangled state for sufficiently generic $\theta_n$. In the above equation, each $|n\rangle$ contains an even number of $1$'s and $0$'s, and together they form a $\mathbb{Z}_2$-symmetric computational basis $\{|n\rangle\}$ with cardinality $N=2^{L-1}$. In this paper, we consider unitaries $U$ composed of local unitary QA gates. With this construction, the entanglement can grow linearly in time, and saturates to volume-law scaling at late times.

Aside from the QA unitary operators, non-unitary local measurements are also introduced into the QA circuit. Since the QA unitary evolution does not enlarge the number of basis states involved in the wave function, repeated local projective measurements in the $Z$ direction will continually reduce the number of available basis states, and will ultimately lead to a product state with no entanglement. Therefore, there is no entanglement phase transition when the measurement rate is finite.

To resolve this issue, Ref.~\onlinecite{Iaconis:2021pmf} introduced a composite measurement which applies a rotation to the spin into $|\pm x\rangle$ following the projection in the $Z$ direction so as to preserve the basis states. In such a hybrid QA circuit model, the wave function at any time is an equal weight superposition of all the basis states, and there exists an entanglement phase transition belonging to DP universality class at finite measurement rate. In our system, we need to modify this composite measurement slightly to preserve the $\mathbb{Z}_2$ symmetry. We therefore define the composite measurement as

\begin{equation}
M_{L/R}^{\sigma}=R\circ P_{L/R}^{\sigma},
\end{equation}
which acts on two qubits. This measurement is a combination of the projection operator  $P_{L/R}^{\sigma}$ on the left/right qubit into the spin $\sigma=\{0,1\}$, together with a two-site rotation operation
\begin{equation}
  R=\frac{1}{\sqrt{2}}\begin{pmatrix}
  1 & 0 & 0 & 1 \\
  0 & 1 & 1 & 0 \\
  0 & 1 & -1 & 0 \\
  1 & 0 & 0 & -1
\end{pmatrix}
 \end{equation} 
that maps $|00\rangle$ to $(|00\rangle+|11\rangle)/\sqrt{2}$, $|11\rangle$ to $(|00\rangle-|11\rangle)/\sqrt{2}$ and $|01\rangle$ to $(|01\rangle+|10\rangle)/\sqrt{2}$, $|10\rangle$ to $(|01\rangle-|10\rangle)/\sqrt{2}$. For instance, when $M_{L}^{0}$ is applied to a two-site wave function with even parity defined as follows,
\begin{equation}\label{Monpsi}
\begin{aligned}
  M_{L}^{0}|\psi\rangle &=R\circ P_{L}^{0}[\frac{1}{\sqrt{2}}(e^{i\theta_0}|00\rangle+e^{i\theta_1}|11\rangle)]
  \\
  &=e^{i\theta_0}R|00\rangle
  \\
  &=\frac{1}{\sqrt{2}}e^{i\theta_0}(|00\rangle+|11\rangle).
\end{aligned}
\end{equation}

After imposing the composite measurement, the wave function is still an equal weight superposition of all the basis states with the same parity: the only thing that changes is the information stored in $|\psi\rangle$, among which only half of the phases are preserved after each application of $M_{L/R}^{\sigma}$. Therefore, we anticipate that measurements will act to disentangle the many-qubit system, while still preserving the $\mathbb{Z}_2$ symmetry.  

\section{Entanglement Transition}\label{Sec: EE}

\subsection{Clifford QA circuit and entanglement dynamics}\label{sec:QA_EE}

  \begin{figure}
  \centering
  \subfigure[]{
    \includegraphics[width=.35\textwidth]{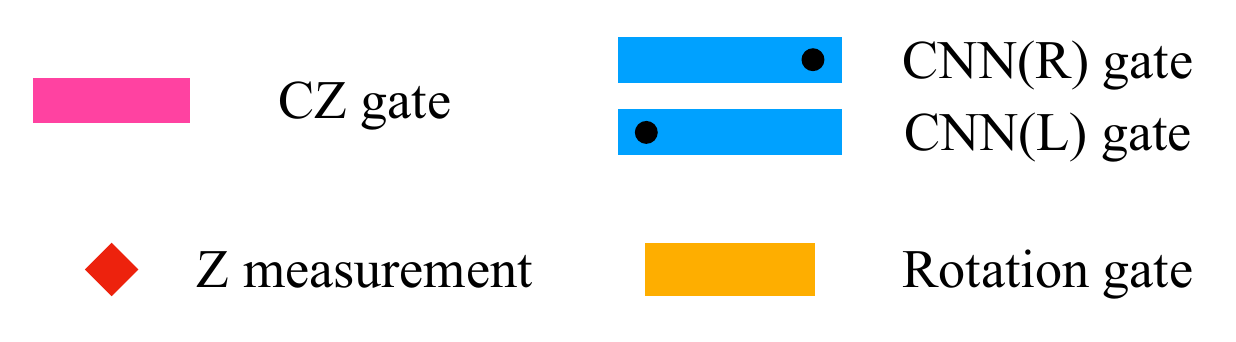}
    \label{fig:EE_gate}}
  \subfigure[]{
    \includegraphics[width=.35\textwidth]{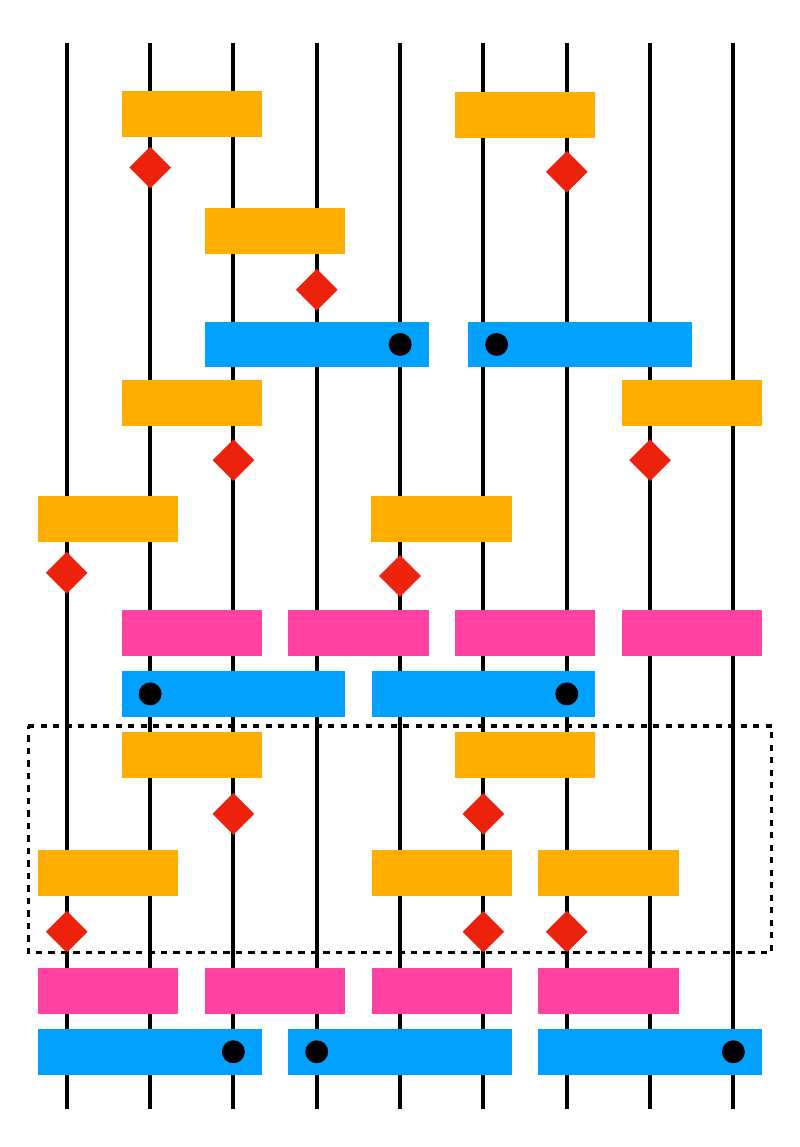}
    \label{fig:QA_EE}}

  \caption{(a) A schematic for the gates appearing in our circuit.  (b)  The arrangement of gates in a single time step of our $\mathbb{Z}_2$-symmetric hybrid QA circuit. Each time step involves three layers of CNN gates and two layers of CZ gates, interspersed with three measured layers. The dashed box represents a measured layer enclosing two rows of composite measurements, with the first/second row containing randomly distributed $M_{L/R}^{\sigma}$ applied on sites $(2i-1,2i)$[$(2i,2i+1)$] for $i\in[1,L/2]$. As with the CNN gates, the projection of $M_{L/R}^{\sigma}$ is chosen to be applied on the left/right qubit with equal probability. In general, the composite measurement appears in a measured layer with probability $p$.}
  \label{fig:entanglement}
  \end{figure}

We choose a subset of Clifford gates to construct a QA circuit with $\mathbb{Z}_2$ symmetry (an example is illustrated in Fig. \ref{fig:entanglement}), and explore the entanglement dynamics by varying the composite measurement rate $p$. First we prepare a product state with $L$ qubits polarized in the $+x$ direction and measure the Pauli string $Z_1Z_2\cdots Z_L$ to implement $\mathbb{Z}_2$ symmetry. We take this as the initial state $|\psi_I\rangle$, and then apply the hybrid circuit, consisting of $\mathbb{Z}_2$-symmetric QA unitaries and composite measurements, to $|\psi_I\rangle$. We then compute the entanglement entropy of a consecutive subsystem $A$. 

Notably, the entanglement dynamics of a Clifford circuit can be efficiently simulated by applying the stabilizer formalism from the Gottesman-Knill theorem \cite{PhysRevA.70.052328}. A stabilizer of a pure state $|\psi\rangle$ is a Pauli string operator $g$ that acts trivially on $|\psi\rangle$, i.e., $g|\psi\rangle=|\psi\rangle$. Such state with $L$ qubits can be uniquely specified by a stabilizer group $G$ generated by $L$ independent and mutually commuting stabilizers,
\begin{equation}
    \begin{aligned}
        G&=\langle \mathcal{G} \rangle=\langle g_1,\dots,g_L\rangle
        \\
        &=\Bigl\{\prod_{i=1}^L g_i^{p_i}|p_i\in \{0,1\}, g_i|\psi\rangle=|\psi\rangle, [g_i,g_j]=0\Bigr\},
    \end{aligned}
\end{equation}
where $\mathcal{G}=\{g_1,\dots,g_L\}$ is the generating set of $G$. By definition, a Clifford unitary gate maps a Pauli string operator to another one, i.e., $UgU^{\dagger}=g', \forall g\in G$. On the other hand, any Pauli measurement $O_i$ acting on the $i$th site becomes a generator of the stabilizer group, with the rest of the generators rearranged so that $O_i$ commutes with all elements in $G$. Consequently, instead of tracing the trajectory of $|\psi\rangle$ with $2^L$ degrees of freedom, we can keep track of the generating set of its stabilizer group whose information can be conveniently stored in a $L\times 2L$ binary matrix. Hence, we are able to perform the simulation on a large system with hundreds of qubits.

The unitary evolution is composed of two types of gates, both of which preserve the $\mathbb{Z}_2$ symmetry. The first type are CNOTNOT(CNN) gates, which are three-qubit gates that flip two qubits according to the value of the third (control) qubit. If the control qubit is on the left we denote the corresponding gate as CNN$_L$; it acts as 
\begin{equation}
\begin{aligned}
  &\text{CNN}_L|1\sigma_1\sigma_2\rangle =|1(1-\sigma_1)(1-\sigma_2)\rangle
  \\
  &\text{CNN}_L|0\sigma_1\sigma_2\rangle=|0\sigma_1\sigma_2\rangle.,
\end{aligned}
\end{equation}
with the leftmost qubit acting as the control. 
The case when the rightmost qubit acts as the control analogously defines a right CNN gate CNN$_R$. In the circuit under consideration,
we choose CNN$_L$ and CNN$_R$ gates randomly, with equal probability. Notice that in each time step, we apply three layers of random CNN gates as shown in Fig.~\ref{fig:QA_EE}.

The second type of gate that appears in the unitary evolution part of the circuit is the CZ gate. This gate is diagonal in the computational basis, and assigns a $\pi$ phase to $|11\rangle$. Explicitly, 
\begin{equation}
\text{CZ}=\begin{pmatrix}
  1 & 0 & 0 & 0 \\
  0 & 1 & 0 & 0 \\
  0 & 0 & 1 & 0 \\
  0 & 0 & 0 & -1
\end{pmatrix}.
\end{equation}
 In the circuit, we apply two layers of CZ gate in each time step.

  \begin{figure*}[ht]
  \centering
  \subfigure[]{
    \includegraphics[width=0.45\textwidth]{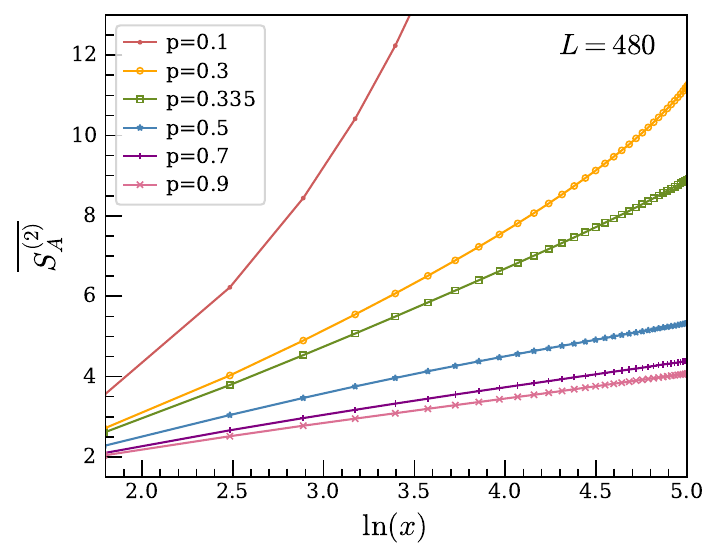}
    \label{fig:EE_LA}}
  \subfigure[]{
    \includegraphics[width=0.45\textwidth]{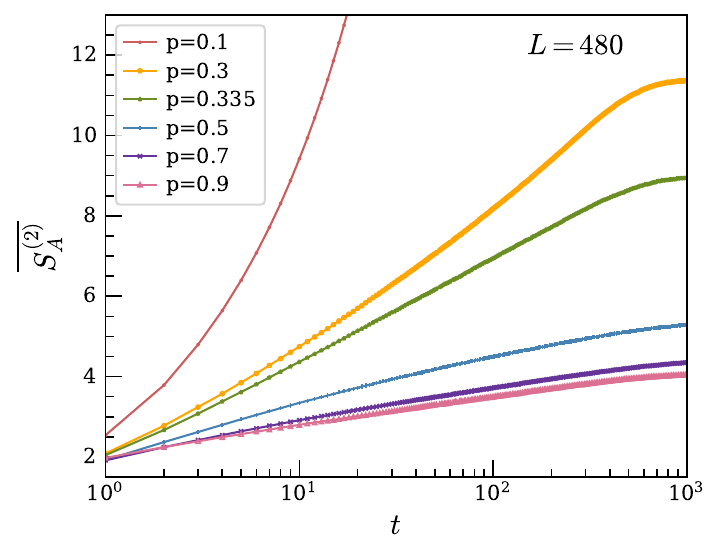}
    \label{fig:EE_t}}

  \subfigure[]{
    \includegraphics[width=0.45\textwidth]{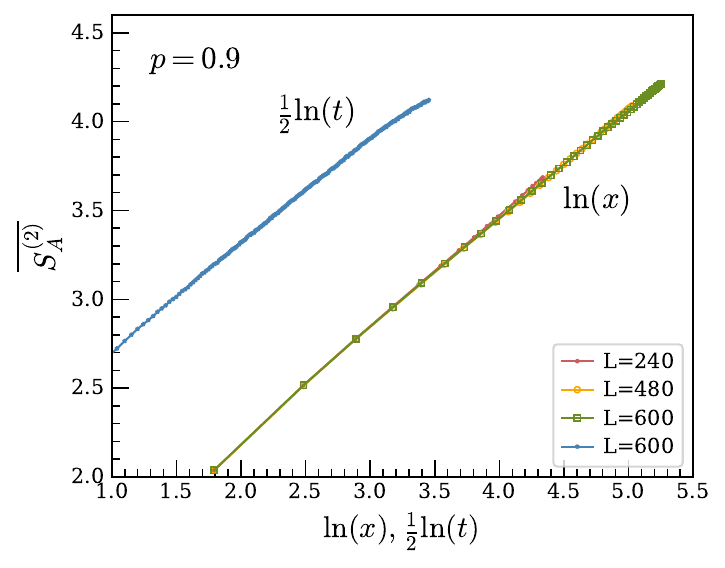}
    \label{fig:EE_dc}}
  \subfigure[]{
    \includegraphics[width=0.45\textwidth]{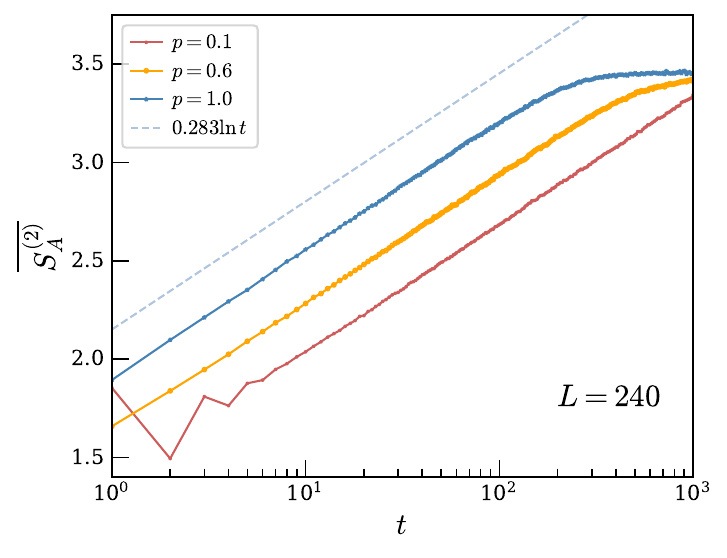}
    \label{fig:EE_noCNN}}
  \caption{(a) The steady state $\overline{S_A^{(2)}}$ vs $\ln{(x)}$ for $L=480$, where $x\equiv \sin{(\pi L_A/L)}L/\pi$. (b) The entanglement dynamics for half of the system $\overline{S_A^{(2)}}$ vs $t$ on the semi-logarithm scale for $L=480$. (c) An example of the data collapse of the steady state $\overline{S_A^{(2)}}$ vs $\ln{x}$ for different system sizes at $p=0.9$. The slope for $L=600$ is $\lambda_2(0.9)=0.605$. We also plot $\overline{S_A^{(2)}}$ vs $\frac{1}{2}\ln{(t)}$ for comparison and we can see that it is roughly parallel to the steady state curves. Numerically, $\lambda_1(0.9)=0.291$. The ratio between these two slopes is $2.079$. On average, $\lambda_2/\lambda_1=2.009$ for $p>p_c$. Similarly, for $p=p_c$, $\lambda_2=1.947$ and $\lambda_1=1.12$, leading to a ratio $\lambda_2/\lambda_1=1.738$. (d) The entanglement dynamics of the QA circuit with no CNN gates for $L=240$ plotted on the semi-log scale. We find that $\overline{S_A^{(2)}(t)}=0.283\ln{(t)}$ for all $p$. All of the numerical data for entanglement entropy are calculated with periodic boundary conditions, and in the natural logarithm base.
}
  \end{figure*}

The randomly-applied composite measurements can be constructed by Clifford gates  defined as $M_{L/R}^{\sigma}$ in Sec.~\ref{sec:QA}. We introduce the composite measurements into the circuit and define the measurement rate $p$ as the density of $M_{L/R}^{\sigma}$ in each measured layer. As we increase $p$ from $0$, the entanglement entropy decreases. Numerically, we observe an entanglement transition at $p_c\approx 0.335$. The value of the critical point is consistent with that observed in the purification dynamics in Sec.~\ref{Sec: PT} and the classical bit-string dynamics in Appendix~\ref{Appendix: BAW}. As shown in Fig.~\ref{fig:EE_LA}, when $p<p_c$, the entanglement entropy has volume-law scaling. The volume-law coefficient decreases as we increase $p$. When $p\geq p_c$, Fig.~\ref{fig:EE_LA} indicates that the steady state entanglement scales logarithmically in the subsystem size. In our numerical simulations, we impose periodic boundary conditions and observe that
\begin{equation}
  \overline{S_A^{(2)}(L_A,p)}=\lambda_2(p)\ln{\Big[\frac{L}{\pi}\sin{(\frac{\pi L_A}{L})}\Big]},
\end{equation}
where the overbar represents an ensemble average. This is interesting and is distinct from conventional measurement-induced phase transitions in interacting systems, where an area-law entangled phase appears for $p>p_c$. In our model, the area-law phase is replaced by a critical phase with $\lambda_2(p)$ changing continuously with $p$. This critical phase is a special feature of the QA circuit with $\mathbb{Z}_2$ symmetry. As we will explain later, this is related to the underlying classical bit-string dynamics with $\mathbb{Z}_2$ symmetry.

Aside from the steady state, we also study the entanglement dynamics starting from an initial state $|\psi_0\rangle$. When $p<p_c$, $S_A(t)$ grows linearly at early times and saturates to a volume-law entangled steady state, while for $p\geq p_c$ we observe a logarithmic entanglement growth before saturation,
\begin{equation}
  \overline{S_A^{(2)}(t,p)}=\lambda_1(p)\ln(t),
\end{equation}
as shown in Fig. \ref{fig:EE_t}. Similar to $\lambda_2(p)$, $\lambda_1(p)$ also depends on $p$. We find that when $p=p_c$, $\lambda_2/\lambda_1=1.738$, while when $p>p_c$ and the circuit is measurement-dominated, the ratio is independent of $p$, with $\lambda_2/\lambda_1=2.009$. 

We also simulate the entanglement dynamics for the QA circuit in the absence of CNN gates. The numerics in Fig.~\ref{fig:EE_noCNN} shows that in such a circuit, the system is critical and has logarithmic entanglement scaling. In particular, $\overline{S_A^{(2)}(t)}=\lambda_1\ln(t)$ where $\lambda_1=0.283$ for all $p$. On the other hand, the steady state entanglement entropy $\overline{S_A^{(2)}}=\lambda_2\ln(x)$ with $\lambda_2=0.591$ for all $p$. Hence the ratio is $\lambda_2/\lambda_1=2.088$ which is close to that in the critical phase of the circuit with CNN gates. In the following sections, we will give an interpretation for $\lambda_1$ and $\lambda_2$ and show that the ratios between them are related to the dynamical exponents of the underlying classical bit-string model. 

\subsection{bit-string dynamics with $\mathbb{Z}_2$ symmetry}\label{sec:bit_EE}

For the second R\'enyi entropy, the purity $\text{Tr}(\rho_A^2)$ is equivalent to the expectation value of the $\mathsf{SWAP}_A$ operator which acts on the tensor product of two identical copies of the state \cite{PhysRevLett.104.157201,islam2015measuring},
\begin{equation}
  \text{Tr}(\rho_A^2)=\langle\psi|_2\otimes\langle\psi|_1 \mathsf{SWAP}_A|\psi\rangle_1\otimes|\psi\rangle_2.
\end{equation}
For the wave function $|\psi\rangle$ expanded in the basis in subregion $A$ and $B$,
\begin{align}
    |\psi\rangle=\frac{1}{\sqrt{N}}\sum_{i,j}e^{i\theta_{ij}}|\alpha_i\rangle_A|\beta_j\rangle_B,
\end{align}
the $\mathsf{SWAP}_A$ operator then exchanges the spin configurations $|\alpha\rangle$ within the $A$ region of the copies of the system (here $N=2^{L-1}$ is the total number of basis states). 

To understand the entanglement dynamics in the non-unitary evolution described by $\tilde{U}_t$, we insert two complete sets of basis states in Eq. \eqref{eq:purity} and find \cite{Iaconis:2021pmf},
\begin{equation}\label{eq:purity}
  \begin{aligned}
    \text{Tr}(\rho_A^2)&=\sum_{n_1,n_2}\langle\psi|_2\langle\psi|_1 \mathsf{SWAP}_A|n_1\rangle|n_2\rangle\langle n_2|\langle n_1|\psi\rangle_1|\psi\rangle_2
    \\
    &=\sum_{n_1,n_2}\langle\psi_0|_2\langle\psi_0|_1\tilde{U}_t^{\dagger}\otimes\tilde{U}_t^{\dagger}|n_1'\rangle|n_2'\rangle
    \\
    & \qquad \qquad \langle n_2|\langle n_1|\tilde{U}_t\otimes\tilde{U}_t|\psi_0\rangle_1|\psi_0\rangle_2
    \\
    &=\frac{1}{N^2}\sum_{n_1,n_2}e^{-i\Theta_{n_1'}}e^{-i\Theta_{n_2'}}e^{i\Theta_{n_1}}e^{i\Theta_{n_2}},
  \end{aligned}
\end{equation}
where 
\begin{equation}
  \begin{aligned}
    |n_1'\rangle|n_2'\rangle&=\mathsf{SWAP}_A|n_1\rangle|n_2\rangle
    \\
    &=\mathsf{SWAP}_A|\alpha_1\beta_1\rangle|\alpha_2\beta_2\rangle
    \\
    &=|\alpha_2\beta_1\rangle|\alpha_1\beta_2\rangle.
  \end{aligned}
\end{equation}
and
\begin{equation}\label{theta_phases}
  e^{i\Theta_n}=\sqrt{N}\langle n|\tilde{U}_t|\psi_0\rangle,\: e^{-i\Theta_n}=\sqrt{N}\langle \psi_0|\tilde{U}_t^{\dagger}|n\rangle.
\end{equation}
The problem of computing $\text{Tr}(\rho_A^2)$ can therefore be converted into evaluating the phases in \eqref{theta_phases}. 

When estimating the overlap of $\tilde{U}_t|\psi_0\rangle$ with any basis state $\langle n|$, we can deduce the effective action of $\tilde{U}_t$ on $\langle n|$ and compute its overlap with $|\psi_0\rangle$ even though the composite measurement is non-unitary. Consider applying a composite measurement $M_{L/R}^{\sigma}$ on $|\psi\rangle$ which is the equal weight superposition of all the allowed states, 

\begin{equation}
  \begin{aligned}
    \langle n|& M_{L/R}^{\sigma}|\psi\rangle =\langle n| R\circ P_{L/R}^\sigma|\psi\rangle
    \\
    &=\langle T_{L/R}^{\sigma}(n)|\psi\rangle = \frac{1}{\sqrt{N}}e^{i\theta_{T_{L/R}^{\sigma}(n)}}.
  \end{aligned}
\end{equation}
Here $|T_{L/R}^{\sigma}(n)\rangle$ refers to the state $|n\rangle$ with the spin at site $L/R$ forced to be in the $\sigma$ state, while its neighboring spin at site $R/L$ is chosen to preserve the parity. Suppose the hybrid QA circuit has the non-unitary dynamics of the form $\tilde{U}_t=M_tU_tM_{t-1}U_{t-1}\cdots$, the overlap can be evaluated by applying $\tilde{U}$ from left to right on $\langle n|$, 
\begin{equation}
  \begin{aligned}
    \langle n|\tilde{U}_t|\psi_0\rangle &=\langle n|M_t U_t M_{t-1} U_{t-1}\cdots|\psi_0\rangle
    \\
    &=\langle T_t(n)|U_t M_{t-1} U_{t-1}\cdots|\psi_0\rangle
    \\
    &=\cdots=\frac{1}{\sqrt{N}}e^{i\Theta_n},
  \end{aligned}
\end{equation}
where $\Theta_n$ is the accumulated phase under time evolution,
\begin{equation}
  e^{i\Theta_{n}}=e^{i\theta_{n(t=1)}} e^{i\theta_{n(t=2)}}\dots e^{i\theta_{n(t=T)}}.
\end{equation}

\begin{figure}
  \centering
  \subfigure[]{
    \includegraphics[width=.35\textwidth]{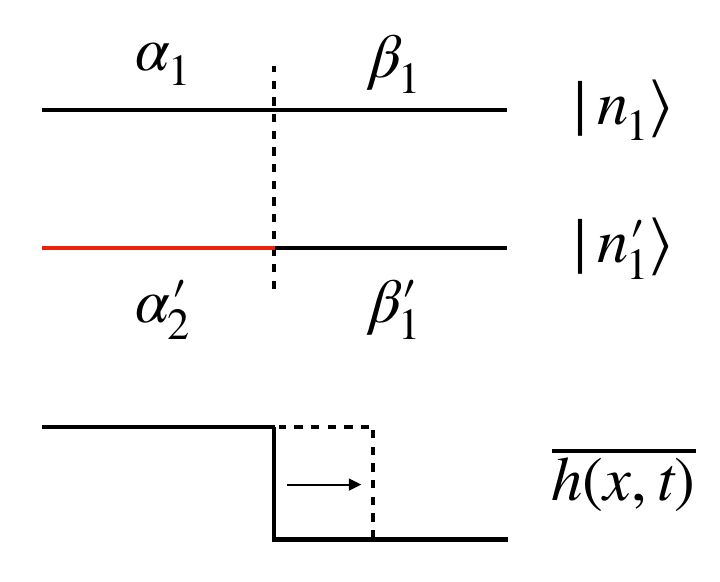}
    \label{fig:bit_diff}
  }
  \subfigure[]{
    \includegraphics[width=.4\textwidth]{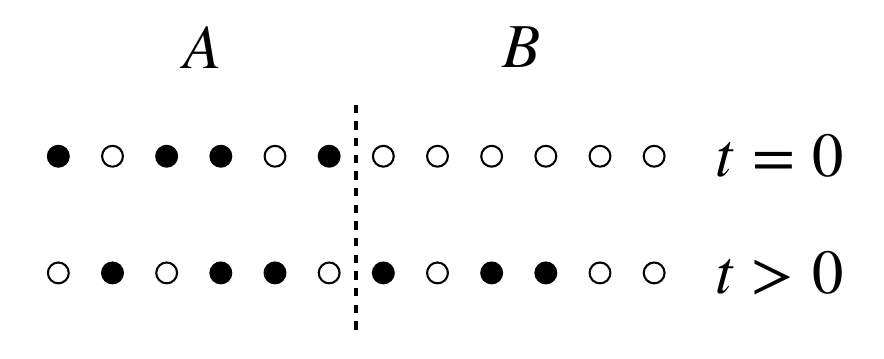}
    \label{fig:particle_rep}
    }
  \caption{(a) The spreading of the bit-string difference $\overline{h(x,t)}$ under the hybrid QA circuit with $\mathbb{Z}_2$ symmetry. Without the intervention of measurements, the front of $\overline{h(x,t)}$ moves to the right at constant velocity with possible broadening. (b) The particle representation of $h(x,t)$. Initially, all the particles are distributed randomly in region A. Under the CNN gates and measurements, the particles perform branching-annihilating random walks and can intrude into region B.}
  \label{fig:bit_part}
\end{figure}

To compute the dynamics of the purity, we investigate the evolution of bit-strings and the associated phases. We define the difference between bit-string pairs as
\begin{equation}
  h(x,t)=|n_1(x,t)-n_1'(x,t)|.
\end{equation}
At each site, $h(x)$ can only be either 0 or 1, and can be conveniently described in terms of the particle representation illustrated in Fig.~\ref{fig:particle_rep} where $\circ$ denotes empty site and $\bullet$ denotes occupied site. For instance, under the CNN$_L$ gate, we have $\bullet\circ\circ\leftrightarrow\bullet\bullet\bullet$ and $\bullet\circ\bullet\leftrightarrow\bullet\bullet\circ$. Under time evolution governed by CNN$_{L/R}$ gates, the particles can diffuse, branch and annihilate on the lattice. Even if the initial configuration only has one particle, the particle number grows linearly in time and the steady state has roughly $L/2$ particles.
On the other hand, under the composite measurement, we have pair-annihilation $\bullet\bullet\to\circ\circ$ and diffusion $\bullet\circ\leftrightarrow\circ\bullet$. The particles diffuse on the lattice and annihilate in pairs with probability $p$ when they encounter one another. Combining unitary dynamics and measurement together, the particles perform branching-annihilating random walks (BAW) with an even number of off-springs \cite{}
\begin{equation}
  W\leftrightarrow 3W, \, W+W \xrightarrow[]{p} \emptyset.
\end{equation}
The competition between the unitary evolution and the composite measurement leads to a continuous phase transition which can be characterized by the total particle number $D(t)\equiv\sum_x h(x,t)$ (The numerical details for this can found in Appendix.\ref{Appendix: BAW}). When $p<p_c$, $D(t\to\infty)/L$ in the steady state saturates to a finite constant. When $p\geq p_c$, if the initial state has an even number of particles, the steady state has $D(t\to\infty)=0$. At $p_c$, $D(t)$ exhibits interesting and universal power law scaling behavior and this critical point belongs to the parity-conserving (PC) universality class with dynamical exponent $z=1.744$\cite{ZHONG1995333,Park_2001,henkel2008non}. When $p>p_c$, the dynamics is dominated by the annihilation process $W+W\to\emptyset$. Since annihilation only occurs when a pair of particles encounter one another, $D(t)$ decays diffusively in time and the $p>p_c$ phase has dynamical exponent $z=2$. This is different from the DP universality class, where a single particle can annihilate directly with probability $p$, which leads to an exponential decay of $D(t)$ with a finite rate at $p>p_c$. The $\mathbb{Z}_2$ symmetry protects the slow diffusive dynamics and is also responsible for the quantum critical phase when we take into account the phase gate. 

Keeping the above classical bit-string dynamics in mind, we  now introduce the phase gate and investigate the entanglement dynamics.
We first consider entanglement entropy for a random phase state defined as
\begin{equation}
  |\psi\rangle=\frac{1}{\sqrt{2^{L-1}}}\sum_n e^{i\theta_n}|n\rangle,
\end{equation}
where $\theta_n$ is a random phase that takes the value in $\{0,\pi\}$ \footnote{In the Clifford dynamics, $\theta_n$ can only take a discrete value $n\pi/2$ with $n$ randomly chosen in 0,1 2 and 3.}. This wave function can be generated under random unitary QA evolution and has maximally entangled volume-law scaling. This can be understood as follows: from Eq. \eqref{eq:purity}, we can see that when $|n_1\rangle=|\alpha_1\beta_1\rangle$ and $|n_2\rangle=|\alpha_2\beta_2\rangle$ share the same spin configuration in region $A$, they are invariant under the swap operator, which means that the random phases always cancel, i.e., $\theta_{n_1}-\theta_{n_1'}=0$ and $\theta_{n_2}-\theta_{n_2'}=0$. There are $2^{L_A}\times (2^{L-L_A-1})^2$ such pairs that each contributes $1/2^{2L-2}$ to the purity. For other bit-strings that are different in region $A$, the random phase terms will in general add up to zero and make no contribution to $\text{Tr}(\rho_A^2)$.\footnote{In fact, the pairs that are the same in region $B$ also contribute to the purity. If we take them into account, the purity becomes $\text{Tr}(\rho_A^2)=(2^{L_A}\times 4^{L-L_A-1}+4^{L_A}\times 2^{L-L_A-1}-2^{L_A}\times 2^{L-L_A-1})/4^{L-1}=2^{-L_A}+2^{-L+L_A}-2^{-L+1}$, therefore the actual steady state entanglement $S_A^{(2)}<L_A\ln{2}$. But now we care about the leading non-constant term so the last two terms are discarded temporarily.} Hence, the wave function has the volume-law scaling
\begin{equation}
  S_A^{(2)}\approx-\ln{\frac{2^{L_A}\times 4^{L-L_A-1}}{4^{L-1}}}=L_A\ln{2}.
\end{equation}

In the above example, only the bit-string pairs without phase difference contribute to the purity. This is also true when we consider the entanglement dynamics starting from $|\psi_I\rangle$. Notice that in Eq. \eqref{eq:purity}, there are four accumulated phases for each bit-string configuration $\{|n_1\rangle,|n_2\rangle,|n_1'\rangle,|n_2'\rangle\}$. We need to find out how these phases evolve in time and how they contribute to the purity. For simplicity, here we first consider the phase difference for $|n_1\rangle$ and $|n_1^\prime\rangle$ only in regime $B$ and define the quantity,
\begin{align}
    Q(t)\equiv\frac{1}{M}\sum_{n_1,n_1^{\prime}} e^{-i\Theta^B_{n_1^\prime}+i\Theta^B_{n_1}},
    \label{eq:Q}
\end{align}
where $M$ is the total number of bit-string pairs. The complete analysis of the time evolution of all these phase terms in the purity will be deferred to Sec.~\ref{sec: double ps}.

Initially, $|n_1\rangle$ and $|n_1^\prime\rangle$ are identical in $B$ and are only different in $A$. The relative phase $-\Theta^B_{n_1^\prime}+\Theta^B_{n_1}$ caused by CZ gates is zero and  we have $Q(t=0)=1$. The nonzero relative phase can be generated when particles enter into $B$. Specifically, if we apply CZ gate on $\bullet  \circ$ with the ensemble of possible bit-string configurations $\{\{|n_1\rangle,|n_1'\rangle\}\}=\{\{|10\rangle,|00\rangle\},\{|11\rangle,|01\rangle\},\{|01\rangle,|11\rangle\},\{|00\rangle,|10\rangle\}\}$, the phase differences generated by the CZ gate are $\{0,\pi,\pi,0\}$. We also get similar results for the particle configuration $\circ \, \bullet$ and $\bullet \, \bullet$. To summarize, for all these nonzero particle configurations, half of the corresponding bit-string pairs contribute a $\pi$ phase to the accumulated phase, while half of them do not contribute any phase terms. This result can be generalized to the many-qubit case. The accumulated phase terms of all the configurations that contain particles in $B$ will add up to zero and make no contribution to Eq. \eqref{eq:Q}. Meanwhile, the configurations that will contribute to $Q(t)$ are those with no particles in $B$ and hence have zero relative phase. Therefore, $Q(t)$ can be alternatively viewed as the fraction of configurations in which the particles never reach the boundary between $A$ and $B$, 
\begin{align}\label{eq:QK}
    Q(t) \approx \frac{K_0(t)}{K},
\end{align}
where $K$ is the total number of particle configurations in $A$ and $K_0$ is the number of particle configurations in which particles never reach the boundary up to time $t$.

\subsection{Single-species BAW model}\label{sec: single ps}
\begin{figure}
  \centering
  \subfigure[]{
    \includegraphics[width=.45\textwidth]{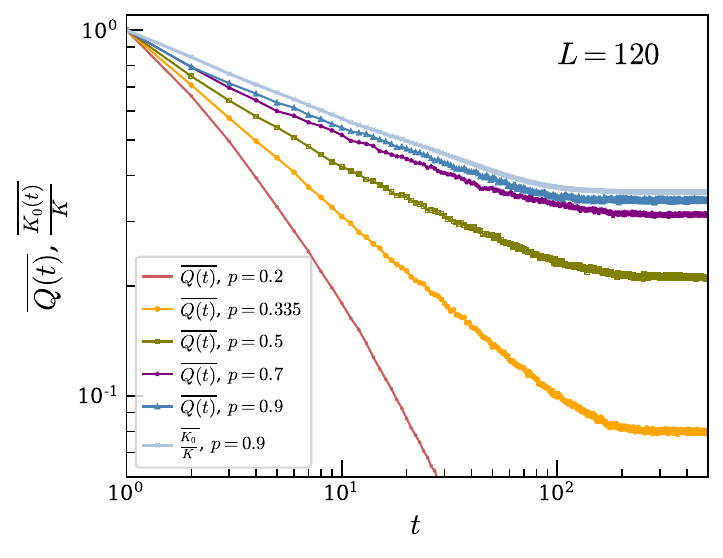}
    \label{fig:bit_1phase}
  }

  \subfigure[]{
    \includegraphics[width=.45\textwidth]{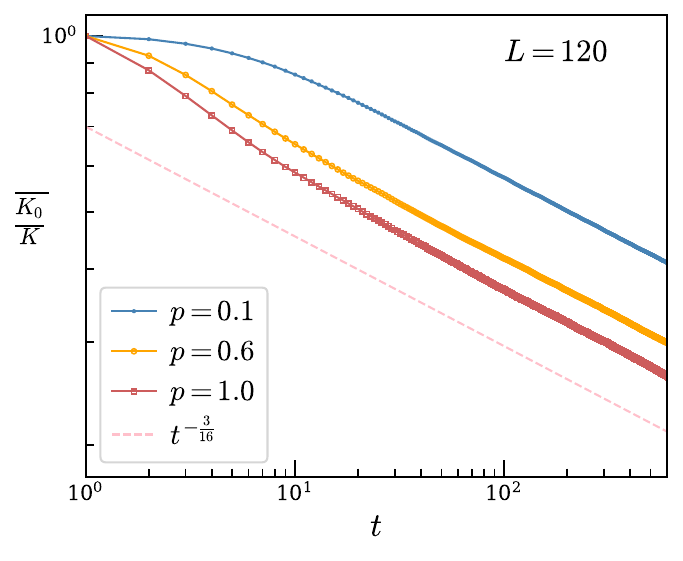}
    \label{fig:ps_noCNN}
  }\textbf{}

  \caption{(a) The evolution of $\overline{Q}$ on a log-log scale. The system size is $L=120$. We also plot $K_0/K$ at $p=0.9$ for comparison. (b) We simulate the single-species BAW model with no CNN gates and plot $\overline{\frac{K_0}{K}}$ vs $t$ for $L=120$ on the log-log scale. $\overline{\frac{K_0}{K}}$ decays as a power law function with the exponent close to the analytical prediction $\frac{3}{16}$.}
 \end{figure}

The above analysis motivates us to define a single-species BAW model. Initially, the particles are distributed randomly in $A$ on a 1D lattice. We let them undergo the same dynamics as the QA circuit in which they perform BAW. Our aim is to find the probability $Q(t)$ that the particles have never reached the boundary between $A$ and $B$ up to time $t$. In the limit where $p=0$, the particle front propagates with a constant velocity $v$. Then, only the initial configurations with no particles distributed within a distance $vt$ to the boundary contribute to $K_0(t)$. This leads to $Q(t)\sim 2^{-vt}$, i.e., the probability that particles never cross the boundary decays exponentially in time. If we roughly take the entanglement entropy as $S_A \sim -\ln Q(t)$, it then grows linearly in time. As we increase $p$, the propagation slows down and eventually becomes diffusion-dominated when $p>p_c$. At this critical point $p_c$ and in the critical phase $p>p_c$, we will see that $Q(t)$ decays algebraically as $Q(t)\sim t^{-\theta}$ where $\theta$ is the so-called persistence exponent in the first passage problem\cite{Bray_2013}. 

We first simulate the phase dynamics and numerically compute $Q(t)$ defined in Eq. \eqref{eq:Q} on an open-boundary 1D lattice in Fig. \ref{fig:bit_1phase}. We find that at $p=p_c$, $\overline{Q(t)}\sim t^{-\theta}$ with $\theta=0.484$ before saturation; when $p>p_c$, $\theta$ decreases by increasing $p$ and the system still stays in the critical phase. We also replace the CZ phase gate by a random phase gate and we observe the same scaling behavior (not presented in the plot). For comparison, we compute the fraction $K_0(t)/K$ and we find that it has the same scaling behavior as $Q(t)$, confirming their equivalence in Eq. \eqref{eq:QK} [See the curves for $p=0.9$ in Fig. \ref{fig:bit_1phase}]. In addition, we also consider the case when there are no CNN gates and the particles only diffuse and annihilate upon contact. As shown in Fig.~\ref{fig:ps_noCNN}, the probability that the particles never cross the boundary scales as $\overline{K_0(t)}/K\sim t^{-3/16}$ for all $p$. The exponent $\frac{3}{16}$ is the persistence rate for the 1D diffusion-annihilation process and has been analytically computed in Refs.~\onlinecite{Derrida1996,Derrida_1995} (For more details, see Appendix.~\ref{Appendix: ps}).

\subsection{Two-species BAW model}\label{sec: double ps}
\begin{figure}
  \centering
  \includegraphics[width=0.45\textwidth]{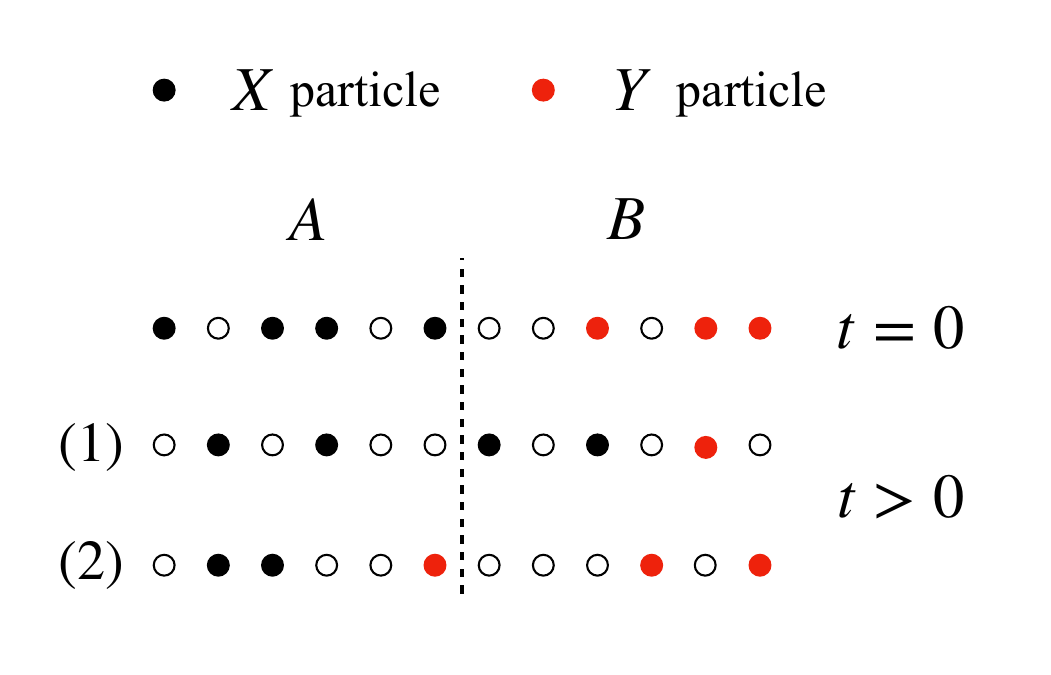}
  \caption{An example of the two-species BAW model. The black dots represent $X$ particles, and the red dots represent $Y$ particles. Initially, $X$ and $Y$ particles are distributed in region $A$ and $B$ respectively. Under the time evolution, the two species perform BAW before they encounter one another. There are two types of possible particle configurations in which the two species have not met up to time $t$: (1) $X$ particles intrude into $B$ and (2) $Y$ particles intrude into $A$.}
  \label{fig:2ps_cartoon}
 \end{figure}

  \begin{figure*}[ht]
  \centering
  \subfigure[]{
    \includegraphics[width=0.45\textwidth]{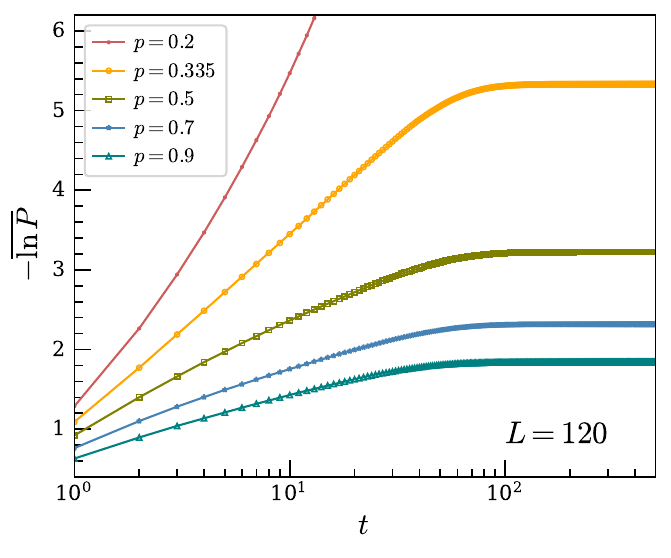}
    \label{fig:2ps_t}
  }
  \subfigure[]{
    \includegraphics[width=0.45\textwidth]{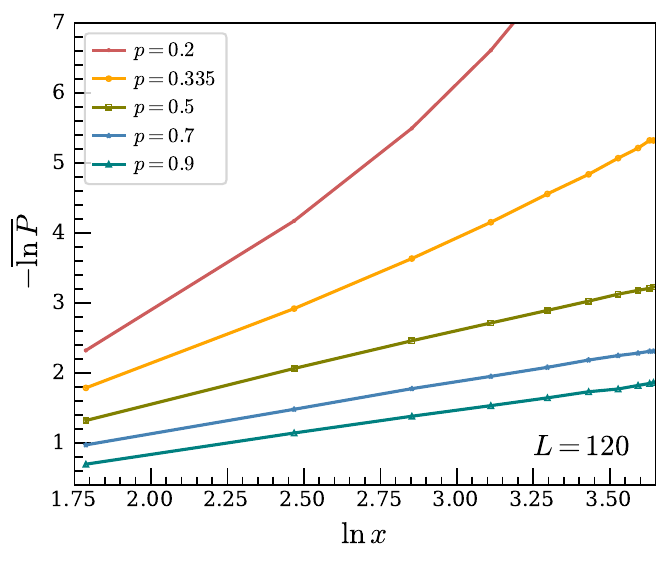}
    \label{fig:2ps_LA}
  }

  \subfigure[]{
    \includegraphics[width=0.45\textwidth]{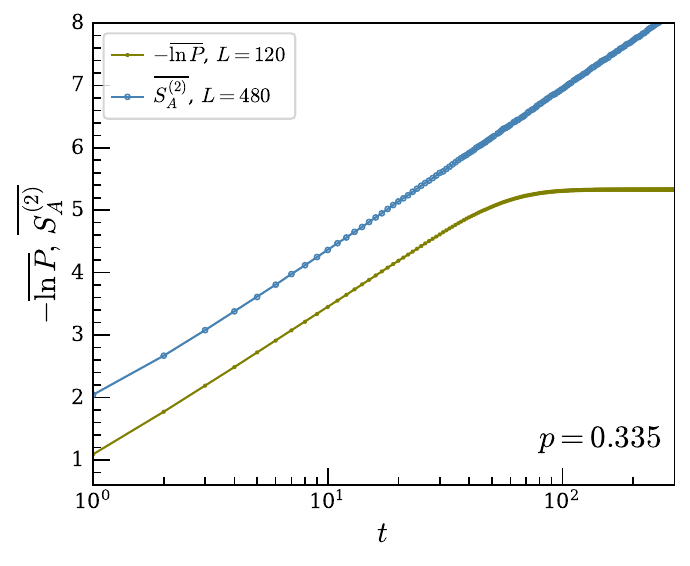}
    \label{fig:2ps_comp}
  }
  \subfigure[]{
    \includegraphics[width=0.45\textwidth]{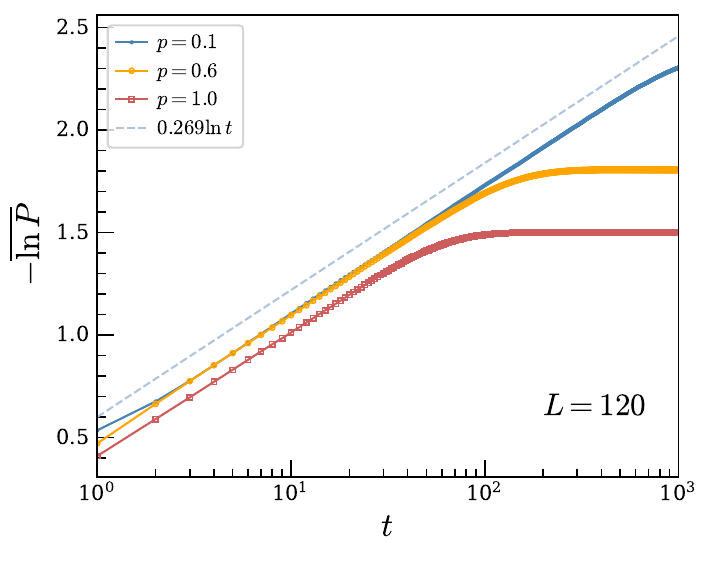}
    \label{fig:2ps_noCNN}
  }

  \caption{(a) $-\overline{\ln{P}}$ vs $t$ on a semi-log scale, defined for a half-system-size cut with system size $L=120$. (b) The steady state $-\overline{\ln{P}}$ vs $\ln(x)$, where $x\equiv \sin{(\pi L_A/L)}L/\pi$. (c) The comparison of $-\overline{\ln{P(t)}}$ and $\overline{S_A^{(2)}(t)}$ at $p=p_c$. (d) The scaling of $-\overline{\ln{P(t)}}$ when the CNN gates are absent. We find that $\lambda_1=0.269$ for all $p$. All of the numerical data of $-\ln{P(t)}$ are calculated under the periodic boundary condition.}
  \label{fig:2ps}
  \end{figure*}
  
Inspired by the single-species BAW model, in this section, we will take into account all of the phase terms and analyze the dynamics of the purity defined in Eq. \eqref{eq:purity}. 

Similar to $Q(t)$ in the previous section, only the bit-string pairs with zero relative phase up to time $t$, viz., those with $ -\theta_{n_1'}-\theta_{n_2'}+\theta_{n_1}+\theta_{n_2}=0$, can contribute to $\text{Tr}[\rho_A^2(t)]$. Any other bit-string pairs will generate random accumulated phase terms, which sum up to zero. 


To understand the zero relative phase constraint, we propose a two-species BAW model. Initially, the particles representing the difference of the bit-string pair $|n_1-n_2|$ are distributed randomly along a 1D lattice. Let $X$ ($Y$) particles denote the bit-string difference initially in region $A$ (region $B$).
We further define $x$ as the location of the rightmost $X$ particle and $y$ as the location of the leftmost $Y$ particle. As shown in Fig.~\ref{fig:2ps_cartoon}, under the hybrid QA circuit with $\mathbb{Z}_2$ symmetry, the particles start to perform BAW. Before $X$ and $Y$ particles encounter one another, the generated phase in each layer $\theta_n$ is composed of three parts: $\theta_n^{[1,x]}$, $\theta_n^{(x,y)}$ and $\theta_n^{[y,L]}$,  which denote the phases generated within the regimes $[1,x]$, $(x,y)$ and $[y,L]$ respectively. Since the first regime occupied by $X$ particles always satisfies $n_1([1,x])=n_2'([1,x])$ and $n_2([1,x])=n_1'([1,x])$, we have $\theta_{n_1}^{[1,x]}=\theta_{n_2^\prime}^{[1,x]}$, $\theta_{n_2}^{[1,x]}=\theta_{n_1^\prime}^{[1,x]}$. Similarly, in the third regime occupied by $Y$ particles,
$\theta_{n_1}^{[y,L]}=\theta_{n_1^\prime}^{[y,L]}$, and $\theta_{n_2}^{[y,L]}=\theta_{n_2^\prime}^{[y,L]}$. In addition, since there is no particle in the intermediate regime, we have $\theta_{n_1}^{(x,y)}=\theta_{n_2}^{(x,y)}=\theta_{n_1^\prime}^{(x,y)}=\theta_{n_2^\prime}^{(x,y)}$. Therefore the total phase difference vanishes: $ -\theta_{n_1'}-\theta_{n_2'}+\theta_{n_1}+\theta_{n_2}=0$.

Once the rightmost $X$ particle runs into the leftmost $Y$ particle, the two-qubit phase gate acting on the $x$th and $y$th sites will generate a random relative phase. Therefore, $\text{Tr}(\rho_A^2)$ is equivalent to the fraction of particle configurations in which two species performing BAW never come across each other,
\begin{equation}
  P(t)=\frac{M_0(t)}{M},
\end{equation}
where $M$ is the total number of particle configurations and $M_0$ is the number of configurations in which $X$ and $Y$ particles never encounter one another up to time $t$.

\begin{table}
\begin{tabular}{|c|c|c|c|c|c|}
    \hline
    \multicolumn{2}{|c|}{ } & $p=0.335$ & $p=0.5$ & $p=0.7$ & $p=0.9$ \\
    \hline
    \multirow{3}{*}{$-\overline{\ln{P}}$} & $\lambda_1$ & 1.053 & 0.507 & 0.355 & 0.293 \\
    & $\lambda_2$ & 1.858 & 0.999 & 0.716 & 0.615 \\
    & $\lambda_2/\lambda_1$ & 1.765 & 1.970 & 2.017 & 2.099 \\
    \hline
    \multirow{3}{*}{$\overline{S_A^{(2)}}$} & $\lambda_1$ & 1.120 & 0.473 & 0.334 & 0.291 \\
    & $\lambda_2$ & 1.947 & 0.926 & 0.665 & 0.605 \\
    & $\lambda_2/\lambda_1$ & 1.738 & 1.958 & 1.991 & 2.079 \\
    \hline
\end{tabular}
\caption{The comparison of scaling prefactors of the two-species BAW model and the $\mathbb{Z}_2$-symmetric Clifford QA model for various measurement rates $p\geq p_c$. Both of them are computed under periodic boundary condition.}
\label{table:2ps_comp}
\end{table}

The validity of the two-species BAW model is numerically verified by simulating $-\ln{P}$ on a $1D$ lattice with periodic boundary condition. Compared with Fig.~\ref{fig:2ps_cartoon},  there are two boundaries between $A$ and $B$. As shown in Fig.~\ref{fig:2ps}, we find that this quantity exhibits a logarithmic growth before saturation, i.e., $-\ln{P(t)}=\lambda_1\ln{t}$ for $p\geq p_c$. Specifically, we compare the value of $-\overline{\ln{P(t)}}$ and $\overline{S_A^{(2)}(t)}$  at $p=p_c$ in Fig.~\ref{fig:2ps_comp} and find that they have the same scaling. Numerically, $\lambda_1(p_c)=1.053\approx 1.12$ where $1.12$ is the prefactor of the logarithmic scaling of $S_A^{(2)}(t)$ at $p=p_c$. In addition, we remove the CNN gates in Fig.~\ref{fig:2ps_noCNN} and let the particles perform diffusion-annihilation random walks. As a result, we find $-\overline{\ln{P(t)}}\sim 0.269\ln(t)$ for all $p$, with the prefactor $0.269$ being close to $0.283$ which is the prefactor of the entanglement entropy without CNN gates.

We also investigate $P$ in the steady state and use this to understand the steady state entanglement entropy. In the steady state, $M_0$ is the number of configurations in which $X$ or $Y$ particles have vanished by annihilating with themselves before they encounter one another. If the subsystem length $L_A\ll L$, it is highly possible that the $X$ particle will vanish first. In this case, when $p\geq p_c$, the subsystem $A$ reaches the steady state at $t\sim L^z_A$ and we have
\begin{align}
    P(t=L_A^z)\sim L_A^{-\lambda_1 z},
\end{align}
this leads to a logarithmic scaling of entanglement entropy with respect to the subsystem length $L_A$. In particular, the prefactor is $\lambda_1 z$.

We simulate $-\ln{P}$ in the steady state in Fig.~\ref{fig:2ps_LA} to numerically verify the above analysis. Here we fix the total system length $L=120$ and vary the subsystem length $L_A$. As expected, we observe a phase transition from the volume-law phase to a critical phase in which 
\begin{align} 
-\ln{P}=\lambda_2\ln{(\sin{(\pi L_A/L)}L/\pi)}
\end{align} 
for $p\geq p_c$. We calculate the ratio between $\lambda_2$ and $\lambda_1$ for different $p$ and find that $\lambda_2/\lambda_1=1.765$ at $p=p_c$ and $\lambda_2/\lambda_1=2.029$ at $p>p_c$, which are consistent with the two dynamical exponents $z=1.744$ at $p=p_c$ and $z=2$ at $p>p_c$ in the PC universality class. These exponents are also very close to the numerical simulation of the Clifford QA model $\lambda_2/\lambda_1=1.738$ at $p_c$ and $\lambda_2/\lambda_1=2.009$ for $p>p_c$. (For a more detailed comparison, see Table. \ref{table:2ps_comp}). Consequently, we can confirm that the hybrid QA model with $\mathbb{Z}_2$ symmetry can be well-described by the classical two-species BAW model.

\section{Purification Dynamics}\label{Sec: PT}
  \begin{figure}[htp!]
  \centering
  \subfigure[]{
    \includegraphics[width=0.35\textwidth]{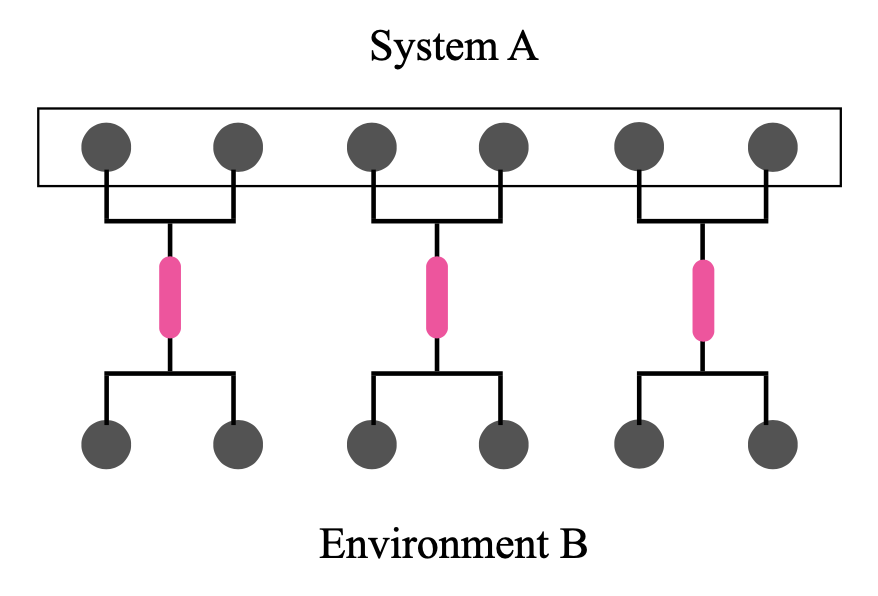}
    \label{fig:Phase_gate}
  }
  \subfigure[]{
    \includegraphics[width=.35\textwidth]{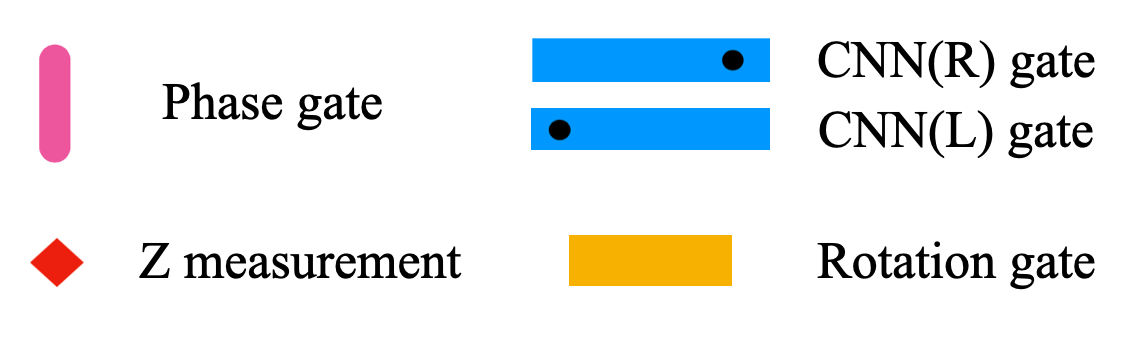}
    \label{fig:Puri_gates}
  }
  \subfigure[]{
        \includegraphics[width=.35\textwidth]{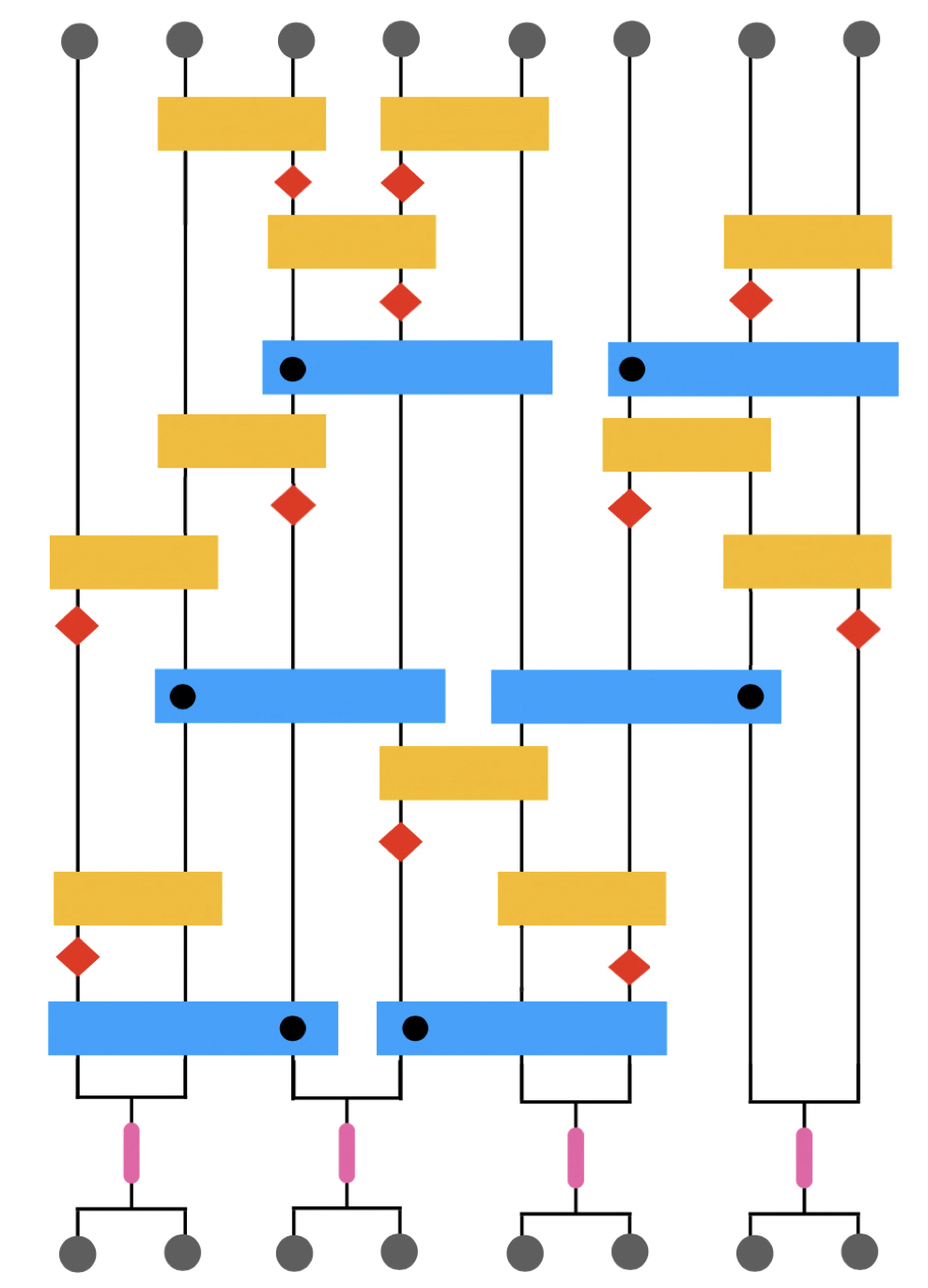}
    \label{fig:QA_puri}
  }

  \caption{Illustration of the circuit used to explore purification dynamics. (a) Every phase gate acts on four qubits, two from system A and two from environment B, in order to form $\frac{L}{2}$ EPR pairs. (b) The symbols of the four-qubit phase gate, three-qubit CNN gate, the single-qubit Z measurement gate and two-qubit rotation gate . (c) The arrangement of gates in a time step for the purification process of $\mathbb{Z}_2$-symmetric hybrid QA circuit model. Except the initial setup in (a), the hybrid circuit is applied in system A only.}
  \label{fig:purification}
  \end{figure}
In this section we will study the purification dynamics of the hybrid QA model with $\mathbb{Z}_2$ symmetry \cite{Gullans_2020}. We consider system $A$ and environment $B$ entangled together, and then apply the hybrid circuit solely on the system $A$. We aim to explore how the entropy of the system depends on the measurement rate.

Under a generic hybrid quantum dynamics, the system will eventually be purified. It is shown in Ref.~\onlinecite{Gullans_2020} that the time of purification can be used to characterize the entanglement phase transition. In the volume-law phase with $p<p_c$, the purification time diverges exponentially in the system size $L$, while in the area-law phase with $p>p_c$, the entropy decays exponentially with a finite rate and the purification time is proportional to $\ln{L}$. At the critical point $p_c$, the entropy decays algebraically when $t\ll L^{z}$. This result also holds in the hybrid QA circuit without $\mathbb{Z}_2$ symmetry, where the purification dynamics can be further interpreted in terms of classical bit-string dynamics \cite{Iaconis:2021pmf}.  

In the presence of the $\mathbb{Z}_2$ symmetry, we will show that the purification dynamics of the QA circuit will be modified when $p>p_c$, analogous to the entanglement dynamics we studied in the previous section. 
Numerically, we prepare a product state with $2L$ qubits polarized in the $x$ direction, and then divide them into system A and environment $B$ with equal size $L$. In order to impose the $\mathbb{Z}_2$ symmetry, we measure the Pauli string $Z_1Z_2\cdots Z_L$ in the system and $Z_{L+1}Z_{L+2}\cdots Z_{2L}$ in the environment. Then we apply a four-qubit diagonal phase gates onto the system A and environment $B$ as in Fig.~\ref{fig:Phase_gate} to create entanglement between them. The phase gate assigns a $\pi$ phase to the basis $|0110\rangle, |0111\rangle, |1110\rangle, |1111\rangle$ with the rest of the basis remaining invariant. Moreover, it is a Clifford gate and therefore the total initial state can be represented as a stabilizer state. Since each phase gate can create $\ln{2}$ entanglement between the system and the environment, the system has an entropy $S_A^{(2)}=\frac{L}{2}\ln{2}$. 

\begin{figure}
  \centering
  \subfigure[]{
    \includegraphics[width=.45\textwidth]{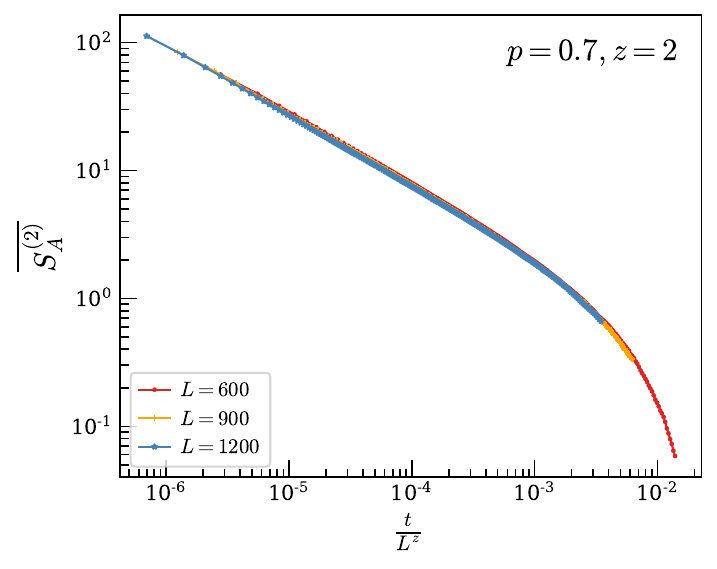}
    \label{fig:puri_dc_2}
  }
  \subfigure[]{
    \includegraphics[width=.45\textwidth]{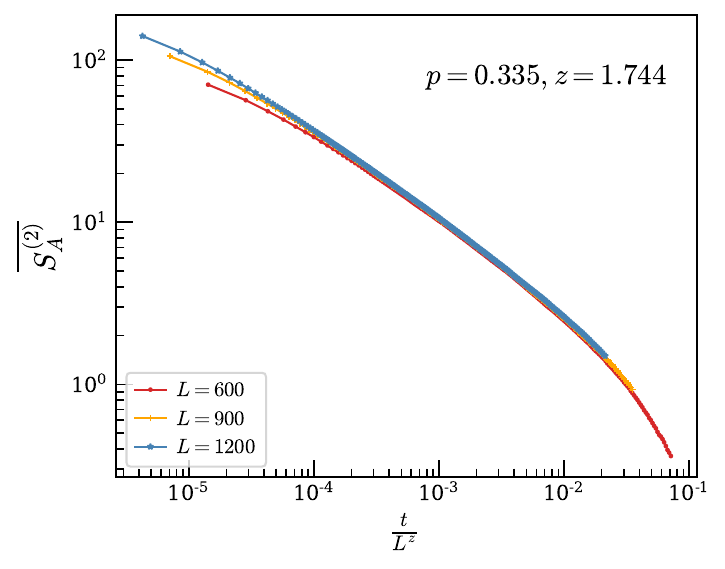}
    \label{fig:puri_dc_1}
  }
  \caption{ Data collapse of purification dynamics described in Fig.~\ref{fig:QA_puri}. (a) is the result at $p=0.7>p_c$ and (b) is the result at $p_c=0.335$.
  }
  \label{fig:puri}
\end{figure}

In the purification dynamics, the unitary and measurement gates are applied solely on system $A$, as shown in Fig.~\ref{fig:QA_puri}. Notice that different from the entanglement process illustrated in Fig.~\ref{fig:QA_EE}, here we do not need to introduce phase gates, due to the fact that the phases between $\{|n_1\rangle,|n_2'\rangle\}$ and between $\{|n_2\rangle,|n_1'\rangle\}$ always cancel with each other. Therefore the unitary evolution consists solely of CNN gates, which simply map one basis state to another. These gates scramble the quantum information within system A, while the entropy of the full system remains the same. On the other hand, the measurement gate disentangles the system from the environment, and the entropy decreases monotonically under the time evolution. 

We simulate the purification dynamics of the above hybrid QA Clifford circuit. When $p> p_c$, we observe that the entropy has a slow diffusive power law decay for a long period of time due to the presence of the $\mathbb{Z}_2$ symmetry, while it takes a time exponentially long in system size to purify the system when $p<p_c$. The data collapse of different system sizes in Fig.~\ref{fig:puri_dc_2} further indicates that $\overline{S_A^{(2)}}=F(t/L^{z})$ with $z=2$ when $p>p_c$. In addition, at critical point $p_c$, the above scaling form also works with different $z=1.744$ [See Fig.~\ref{fig:puri_dc_1}]. We believe that such scaling is universal in other non-Clifford hybrid QA circuits with $\mathbb{Z}_2$ symmetry and the dynamical exponents are consistent with what we found in the entanglement dynamics.

\section{Conclusion}
\label{sec:concl}
In this paper, we explore the $\mathbb{Z}_2$-symmetric quantum automaton (QA) circuit subject to local composite measurements. By tuning the measurement rate $p$, we find an entanglement phase transition from a volume-law entangled phase to a critical phase with logarithmic entanglement scaling. By analyzing the underlying classical bit-string dynamics, we show that the critical point $p_c$ belongs to the parity-conserving universality class. We further show that the critical phase is protected by the combination of $\mathbb{Z}_2$ symmetry and the special feature of QA circuit. We derive an effective two-species particle model in which particles perform branching-annihilating random walks. We use this model to understand the entanglement dynamics and illustrate that the purity of the wave function is equivalent to the fraction of particle configurations in which two different species of particles never encounter. Based on this result, we show that the prefactors of the logarithmic second R\'enyi entropy at the critical point and the critical phase are related to the local persistence exponents of the corresponding two-species particle models. In addition, the above critical behavior when $p\geq p_c$ is further demonstrated in the purification process.

The idea of presenting bit-string dynamics in the particle language can also be applied in Ref.~\onlinecite{Iaconis:2021pmf} to explain the entanglement phase transition without $\mathbb{Z}_2$ symmetry that belongs to the directed percolation universality class. Based on this method, it is also possible to develop similar tools to understand the universality classes of entanglement phase transition in the hybrid Haar random circuit and hybrid Clifford random circuit\cite{skinner2019measurement,Li_2018}. In addition, it can also be used to understand the subleading correction term in the non-thermal volume-law phase when $p<p_c$\cite{fan2020self,li2020statistical}. We leave these interesting questions for future study. 

\acknowledgements
We acknowledge Ethan Lake for his careful proofreading. We also acknowledge the helpful discussions with Jason Iaconis. 

\appendix
\section{parity-conserving universality class and the branching-annihilating random walks}\label{Appendix: BAW}
Nonequilibrium phase transitions in classical dynamical lattice models can be classified purely by their scaling behavior. The most common nonequilibrium class is the directed percolation (DP) universality class. Another class called parity-conserving (PC) universality class emerges when we add extra symmetry, namely, parity conservation to the system. Like the DP universality class, the PC universality class is very robust in a sense that it contains many models that share the same critical exponents. In this appendix, we will show that the BAW model introduced in Sec.~\ref{Sec: EE} belongs to the PC universality class.

\begin{figure*}
  \centering
  \subfigure[]{
    \includegraphics[width=0.45\textwidth]{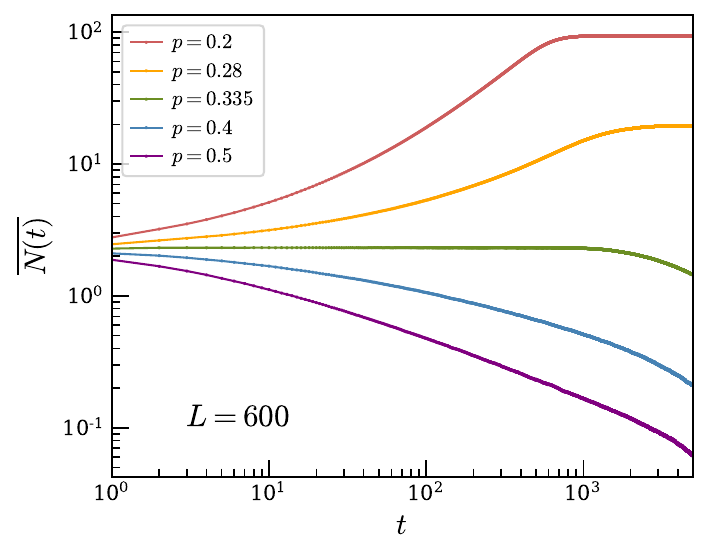}
    \label{fig:PsdN}
  }%
  \subfigure[]{
    \includegraphics[width=0.45\textwidth]{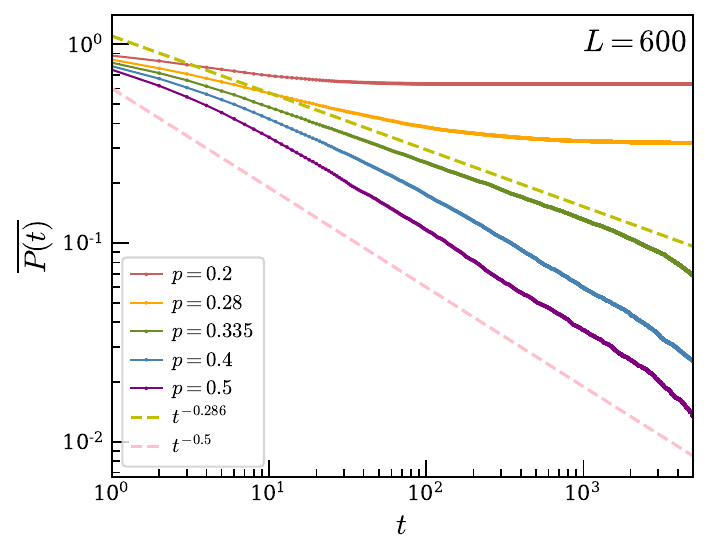}
    \label{fig:PsdP}
  }

  \subfigure[]{
    \centering
    \includegraphics[width=0.45\textwidth]{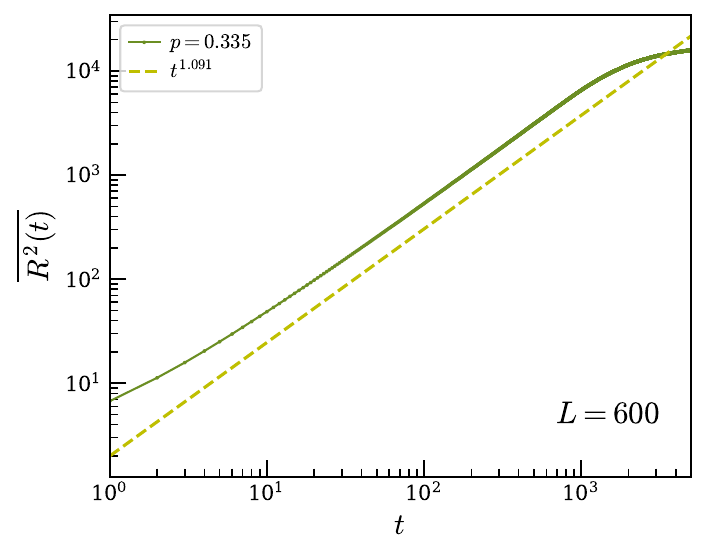}
    \label{fig:PsdMSD}
  }
    
  \caption{We simulate the BAW model of the seeding process starting with a pair of adjacent particles and find that the critical point is around $p_c=0.335$. In the calculation done in the Appendix with $L=600$, we find that if we choose $p_c=0.335$, the critical exponents have the best match with the critical exponents of the PC universality class. (a) The mean particle number $\overline{N}$ vs $t$ on the log-log scale for $L=600$. (b) ${P}$ vs $t$ on a log-log scale for $L=600$. When $p=p_c$, $P(t)\sim t^{-0.286}$ and when $p>p_c$, $P(t)\sim t^{-0.5}$. (c) The mean-square distance scales as $\overline{R^2(t)}\sim t^{1.091}$ at $p=p_c$ for $L=600$.}
  \label{fig:Psd}
\end{figure*}

In Sec.~\ref{Sec: EE} we have established the connection between the hybrid QA model with $\mathbb{Z}_2$ symmetry and a classical particle model. Under the QA circuit composed of CNN gates and composite measurements, the particles perform the branching-annihilating random walks (BAW) where they diffuse on a one-dimensional lattice and annihilate when they come into contact with probability $p$. Furthermore, each particle can generate an even number of off-springs, $i.e.$
\begin{equation}
  W \leftrightarrow 3W,\: W+W\xrightarrow[]{p}\emptyset.
\end{equation}
There are three initial conditions which lead to different scaling behavior of various properties under the same dynamics: (a) the seeding process starting with a pair of adjacent particles, (b) the seeding process starting with a single particle, and (c) the purification process starting with a fully occupied state.

We first analyze the BAW model with initial condition (a) numerically. We vary $p$ and measure the scaling behavior of the mean particle number $\overline{N(t)}$. As shown in Fig.~\ref{fig:PsdN}, we observe a phase transition while adjusting $p$: when $p<p_c\approx0.335$, an active steady state with finite number of particles emerges. At $p=p_c$, $\overline{N(t)}\sim t^{\theta}$ where $\theta=0$. When $p>p_c$, the dynamics is dominated by annihilation of particles in pairs and the system enters an absorbing phase where the particle number is monotonically decreasing until $\overline{N(t\to\infty)}=0$. In addition, we measure two other quantities: $P(t)$, the probability that the system has not entered the absorbing phase at time $t$; $\overline{R^2(t)}$, the mean-square distance from the center of the lattice chain, averaged over the surviving samples. From Fig.~\ref{fig:PsdP}, when $p<p_c$, the system maintains a finite possibility to survive and stay away from the absorbing phase. When $p=p_c$, $P(t)\sim t^{-\delta}$ where $\delta=0.286$. Notably, when $p>p_c$, $P(t)$ still decays as a power law with the exponent $1/z=1/2$. $P$ can also be viewed as an order parameter which marks the existence of a phase transition. Furthermore, the numerics in Fig.~\ref{fig:PsdMSD} shows that the mean-square distance $\overline{R^2(t)} \sim t^{2/z}$ at $p=p_c$ with the other dynamical exponent $z=1.833$. These exponents are universal for the PC universality class and agree with the numerical findings that $\delta=0.286$, $\theta=0$, $z=1.744$ when $p=p_c$ and $z=2$ for $p>p_c$ in Ref.~\onlinecite{ZHONG1995333}.

\begin{figure}
  \subfigure[]{
    \includegraphics[width=0.45\textwidth]{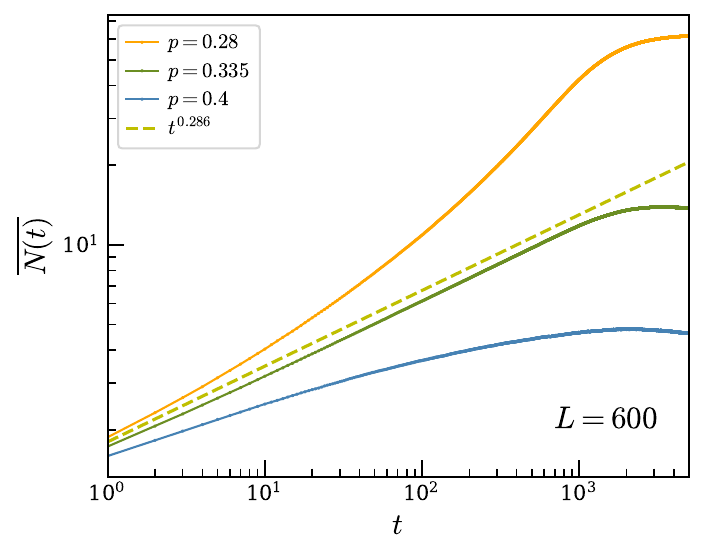}
    \label{fig:PssN}
  }

  \subfigure[]{
    \includegraphics[width=0.45\textwidth]{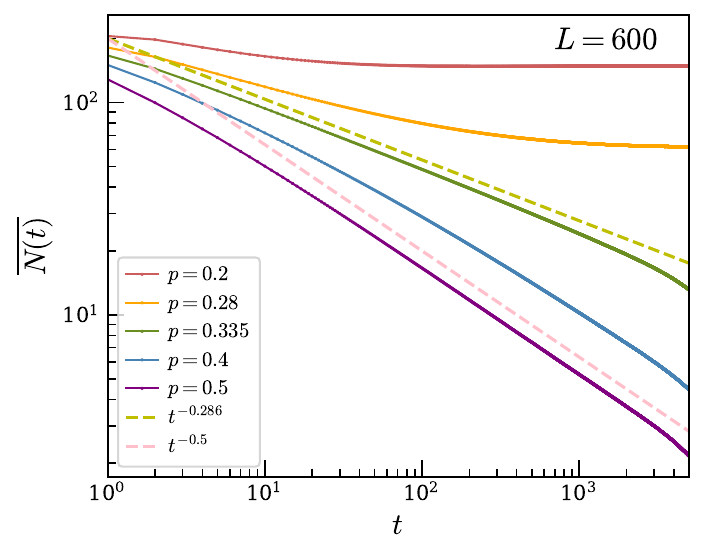}
    \label{fig:PPN}
  }

  \caption{The mean particle number $\overline{N(t)}$ vs $t$ on a log-log scale for (a) the seeding process beginning with a single particle and (b) the purification process starting with a fully occupied state.}
\end{figure}

We also study the other initial conditions under the same dynamics. Fig.~\ref{fig:PssN} exhibits the scaling of $\overline{N(t)}$ for the seeding process starting with a single particle. It is easy to see that the system will never reach an empty state for $N(0)=1$ since the parity is conserved, therefore, the survival rate $P(t)$ is always zero, $\delta=0$ for all $p$. On the other hand, $\overline{N(t)}\sim t^{0.286}$ when $p=p_c$. These exponents coincide with that of the seeding process starting with a pair of particles except that the values of $\delta$ and $\theta$ exchange, which is quite interesting.

As shown in Fig.~\ref{fig:PPN}, $\overline{N(t)}$ for the purification process has a similar scaling with $P(t)$ for the seeding process starting with a pair of adjacent particles. When the measurement rate $p<p_c$, the system approaches an active state with a finite number of particles. Once $p= p_c$, $\overline{N(t)}\sim t^{-0.286}$. When $p>p_c$, the particles are performing annihilation-dominated BAW, $\overline{N(t)}$ still decays algebraically, i.e., $\overline{N(t)}\sim t^{-1/2}$.


\section{Single-species BAW model and the first passage problem}\label{Appendix: ps}
\begin{figure}
  \centering
  \subfigure[]{
    \includegraphics[width=0.3\textwidth]{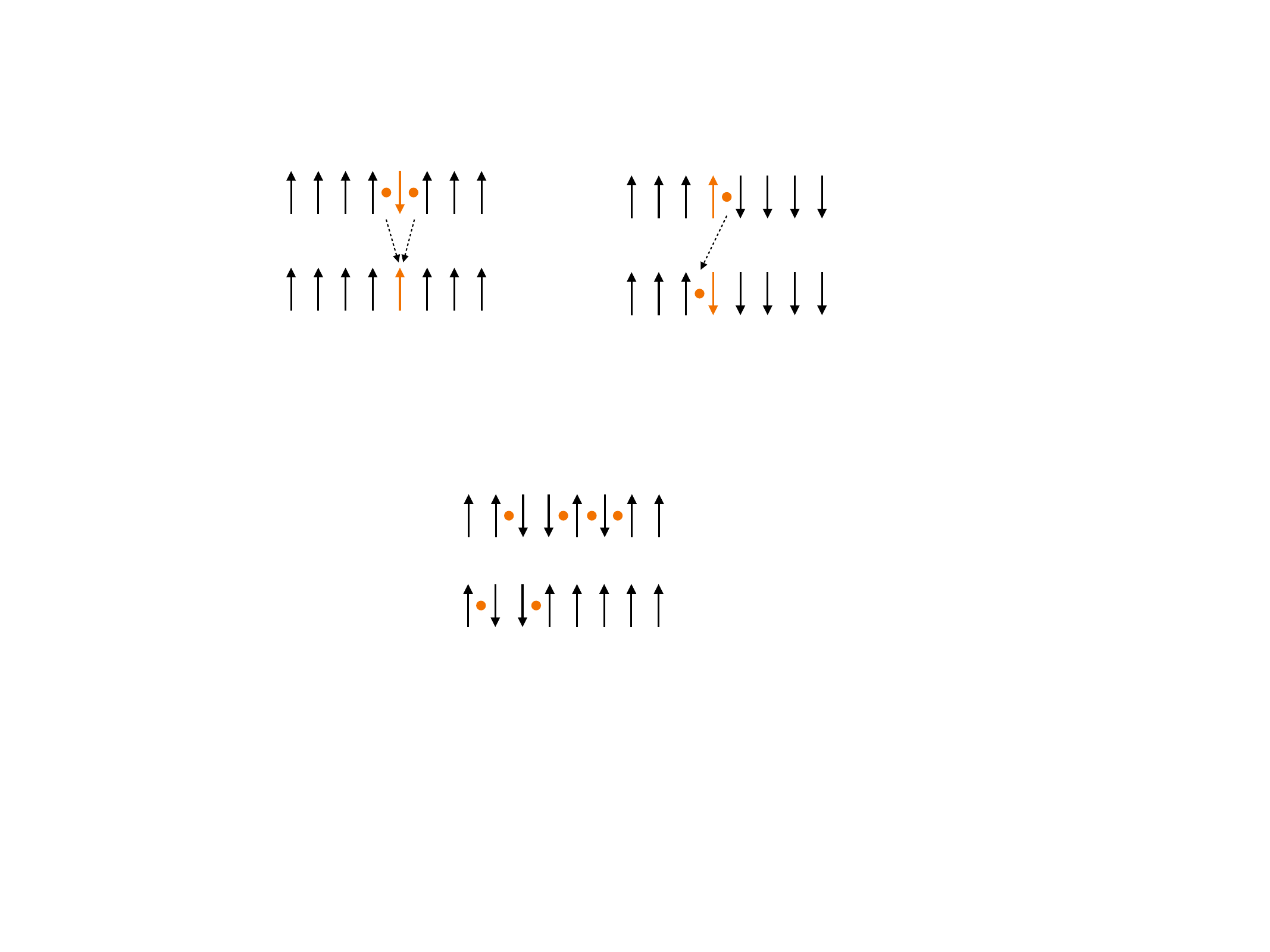}
  }

  \subfigure[]{
    \includegraphics[width=0.3\textwidth]{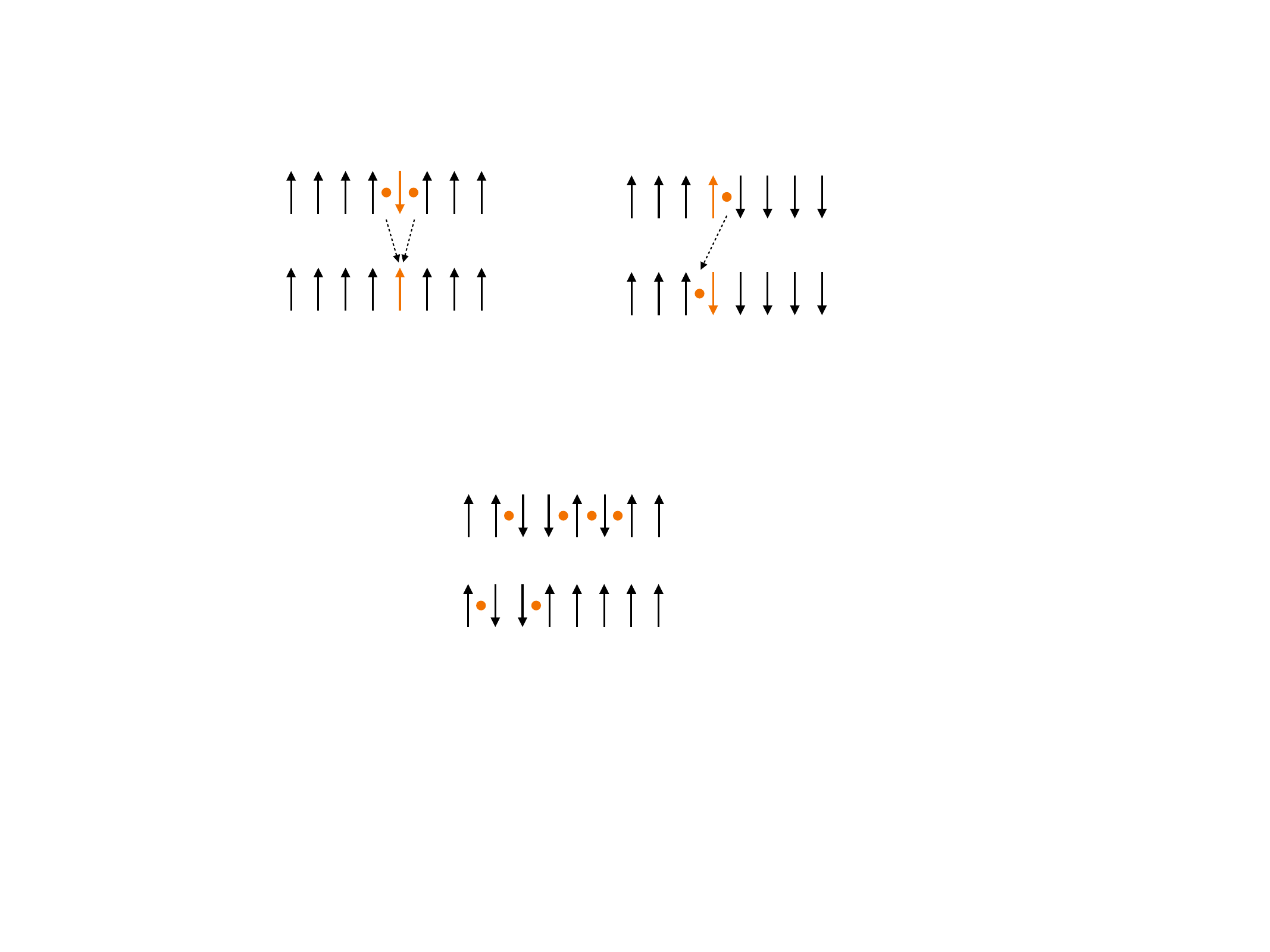}
  } 

  \caption{Mapping between the zero-temperature Glauber dynamics in one dimension and the corresponding domain wall quasiparticles: (a) the spin marked in orange is updated and the domain wall quasiparticles annihilate; (b) the spin marked in orange is flipped to match the value of its right neighbor and the domain wall quasiparticle diffuses to the left.}
  \label{fig:glauber}
\end{figure}

In this appendix, we will investigate the correspondence between the single-species BAW model in Sec. \ref{sec: single ps} and the first passage problem of the 1D Ising model discussed in Ref.~\onlinecite{Derrida1996}. 

In Ref.~\onlinecite{Derrida1996}, they studied the persistence probability $r(q,t)$ that a given spin stays in the same state up to time $t$ of an infinite 1D $q$-state Potts model whose update rule obeys the zero-temperature Glauber dynamics. If a random initial $q$-state spin configuration is quenched at zero temperature, the dynamics tends to align all the spins. At each time step, a chosen spin is updated according to the values of its two nearest neighbors, i.e., $S_i(t+1)=S_{i-1}(t)$ or $S_{i+1}(t)$ with equal probability. They proposed a coagulation model which treats $S_0(t)$ at different time steps as random walkers which coalesce upon contact in the time-reversed order and find that the persistence rate is just the probability that $S_0(1)=S_0(2)=\cdots=S_0(t)$ which scales as
\begin{equation}
  r(q,t)\sim t^{-\theta(q)},
\end{equation}
where the exponent has the analytical expression
\begin{equation}
  \theta(q)=-\frac{1}{8}+\frac{2}{\pi^2}\left[\cos^{-1}\left(\frac{2-q}{\sqrt{2}q}\right)\right]^2.
\end{equation}

A single-species BAW model was introduced in Sec.~\ref{sec: single ps}. Initially, the particles are distributed randomly in the left half of the lattice chain. Under the unitary gates and composite measurements, the particles perform BAW. We have demonstrated that $Q(t)$ defined in Eq.~\ref{eq:Q} is equivalent to $-\ln(K_0/K)$, where $K_0(t)/K$ is the fraction of particle configurations in which the particles never diffuse into the right half of the lattice chain up to time $t$, or in other words, the probability that the boundary between A and $B$ has never been visited by the particles. If we consider the case when the particles are performing diffusion-annihilation random walks, i.e., there are no CNN gates, and we treat them as domain walls between the spins, then their dynamics under the measurement-only circuit has a one-to-one correspondence to the zero-temperature Glauber dynamics of the 1D Ising model $(q=2)$. As illustrated in Fig.~\ref{fig:glauber}, when the spin different from both of its nearest neighbors is flipped, the domain walls annihilate; When its neighbors are in different states and the spin is aligned with one of them, the domain wall either diffuses or stays still. Besides, there is no creation of domain walls, i.e., no particle branching since the zero temperature prohibits any energy-raising move. Then $K_0(t)/K=\sqrt{r(q=2,t)}$, since it is equivalent to the probability that the spin on the boundary of a finite chain has never flipped. Thus, $K_0(t)/K$ decays as a power law with the exponent $\theta(q=2)/2=3/16$.

  \bibliographystyle{apsrev}
  \bibliography{reference}

\end{document}